\documentclass[11pt]{xart}
\usepackage{epsfig}
\begin{document}
\setcounter{page}{1}
\setcounter{tocdepth}{2}
\font\sixrm=cmr6
\font\eightrm=cmr8
\abovedisplayskip=12pt plus 3pt minus 3pt
\belowdisplayskip=12pt plus 3pt minus 3pt
\renewcommand{\thefootnote}{\fnsymbol{footnote}}
\def\thisyear{\number\year}
\def\thismonth{\ifcase\month\or
January\or February\or March\or April\or May\or June\or
July\or August\or September\or October\or November\or December\fi}
\def\Bibitem#1{\bibitem{#1}\label{#1}}
\def\Cite#1{\ref{#1}}
\def\exercise#1{\vspace{2ex}\noindent{\small\sc Exercise:}~{\small#1}
\par\vspace{2ex}}
%\def\exercise#1{}

% Definitions and abbreviations

% Roman letters in math formulae

\def\rmd{{\rm d}}
\def\rmD{{\rm D}}
\def\rme{{\rm e}}
\def\rmO{{\rm O}}

% Special relations and symbols

\def\defeq{\mathrel{\mathop=^{\rm def}}}
\def\proof{\noindent{\sl Proof:}\kern0.6em}
\def\endproof{\hskip0.6em plus0.1em minus0.1em
\setbox0=\null\ht0=5.4pt\dp0=1pt\wd0=5.3pt
\vbox{\hrule height0.8pt
\hbox{\vrule width0.8pt\box0\vrule width0.8pt}
\hrule height0.8pt}}
\def\frac#1#2{\hbox{$#1\over#2$}}
\def\dual{\mathstrut^*\kern-0.1em}
\def\mod{\;\hbox{\rm mod}\;}
\def\ring{\mathaccent"7017}
\def\lvec#1{\setbox0=\hbox{$#1$}
    \setbox1=\hbox{$\scriptstyle\leftarrow$}
    #1\kern-\wd0\smash{
    \raise\ht0\hbox{$\raise1pt\hbox{$\scriptstyle\leftarrow$}$}}
    \kern-\wd1\kern\wd0}
\def\rvec#1{\setbox0=\hbox{$#1$}
    \setbox1=\hbox{$\scriptstyle\rightarrow$}
    #1\kern-\wd0\smash{
    \raise\ht0\hbox{$\raise1pt\hbox{$\scriptstyle\rightarrow$}$}}
    \kern-\wd1\kern\wd0}

% Lattice derivatives

\def\nab#1{{\nabla_{#1}}}
\def\nabstar#1{\nabla\kern-0.5pt\smash{\raise 4.5pt\hbox{$\ast$}}
               \kern-4.5pt_{#1}}
\def\drv#1{{\partial_{#1}}}
\def\drvstar#1{\partial\kern-0.5pt\smash{\raise 4.5pt\hbox{$\ast$}}
               \kern-5.0pt_{#1}}
\def\drvtilde#1{{\tilde\partial_{#1}}}

% Units

\def\MeV{{\rm MeV}}
\def\GeV{{\rm GeV}}
\def\TeV{{\rm TeV}}
\def\fm{{\rm fm}}

% Constants

\def\euler{\gamma_{\rm E}}

% Fields

\def\Nf{N_{\rm f}}
\def\psibar{\overline{\psi}}
\def\phieff{\phi_{\rm eff}}
\def\phiimpr{\phi_{\kern0.5pt\hbox{\sixrm I}}}
\def\phir{\phi_{\hbox{\sixrm R}}}
\def\ar{A_{\hbox{\sixrm R}}}
\def\pr{P_{\hbox{\sixrm R}}}
\def\pMSbar{P_{\hskip0.5pt\lower1pt\hbox{$\scriptstyle\smallmsbar$}}}
\def\Or{{\cal O}_{\hbox{\sixrm R}}}
\def\sr{S_{\hbox{\sixrm R}}}
\def\vr{V_{\hbox{\sixrm R}}}
\def\aimpr{A_{\hbox{\sixrm I}}}
\def\pimpr{P_{\hbox{\sixrm I}}}
\def\vimpr{V_{\hbox{\sixrm I}}}
\def\op#1{{\cal O}_{ #1}}
\def\ophat#1{\widehat{\cal O}_{ #1}}
\def\rhoprime{\rho\kern1pt'}
\def\rhobar{\bar{\rho}}
\def\rhobarprime{\rhobar\kern1pt'}
\def\rhobartilde{\kern2pt\tilde{\kern-2pt\rhobar}}
\def\rhobartildeprime{\kern2pt\tilde{\kern-2pt\rhobar}\kern1pt'}

% Dirac matrices

\def\dirac#1{\gamma_{#1}}
\def\diracstar#1#2{
    \setbox0=\hbox{$\gamma$}\setbox1=\hbox{$\gamma_{#1}$}
    \gamma_{#1}\kern-\wd1\kern\wd0
    \smash{\raise4.5pt\hbox{$\scriptstyle#2$}}}

% Improvement coefficients

\def\ba{b_{\rm A}}
\def\bp{b_{\rm P}}
\def\bg{b_{\rm g}}
\def\bm{b_{\rm m}}

\def\ca{c_{\rm A}}
\def\csw{c_{\rm sw}}

% Correlation functions

\def\fa{f_{\rm A}}
\def\fda{f_{\delta{\rm A}}}
\def\fp{f_{\rm P}}

% Gauge group

\def\SUtwo{{\rm SU(2)}}
\def\SUthree{{\rm SU(3)}}
\def\SUn{{\rm SU}(N)}
\def\SU{{\rm SU}}
\def\tr{\,\hbox{tr}\,}
\def\Ad{{\rm Ad}\,}
\def\CF{C_{\rm F}}

% Chiral transformations

\def\da{\delta_{\hbox{\sixrm A}}}
\def\dv{\delta_{\hbox{\sixrm V}}}

% Action

\def\Sg{S_{\rm G}}
\def\Sf{S_{\rm F}}
\def\Seff{S_{\rm eff}}
\def\Simpr{S_{\rm impr}}
\def\Zf{{\cal Z}_{\rm F}}
\def\Zimprf{\tilde{\cal Z}_{\rm F}}

% Renormalization schemes

\def\alphaSF{\alpha_{\rm SF}}
\def\alphaMSbar{\alpha_{\lower1pt\hbox{$\scriptstyle\smallmsbar$}}}
\def\alphaqqbar{\alpha_{\rm q\bar{q}}}
\def\alphaP{\alpha_{\hbox{\sixrm P}}}
\def\Fqqbar{F_{\rm q\bar{q}}}
\def\LambdaMSbar{\Lambda_{\smallmsbar}}

\def\gmsbar{\bar{g}_{\kern0.5pt\smallmsbar}}
\def\gbar{\bar{g}}
\def\gbarMS{\gbar_{\ms}}
\def\gbarMSbar{\gbar_{\msbar}}
\def\gbarSF{\gbar_{\rm SF}}
\def\gr{g_{{\hbox{\sixrm R}}}}
\def\glat{g_{\lat}}
\def\gSF{g_{{\hbox{\sixrm SF}}}}

\def\mq{m_{\rm q}}
\def\mqtilde{\widetilde{m}_{\rm q}}
\def\mr{m_{{\hbox{\sixrm R}}}}
\def\mc{m_{\rm c}}
\def\mp{m_{\rm p}}
\def\mlat{m_{\lat}}
\def\mSF{m_{\hbox{\sixrm SF}}}
\def\mbar{\overline{m\kern-1pt}\kern1pt}

\def\za{Z_{\rm A}}
\def\zv{Z_{\rm V}}
\def\zp{Z_{\rm P}}
\def\xp{X_{\rm P}}
\def\zpmom{\zp^{\raise1pt\hbox{\sixrm MOM}}}
\def\xpmom{\xp^{\raise1pt\hbox{\sixrm MOM}}}
\def\ztwomom{Z_2^{\raise1pt\hbox{\sixrm MOM}}}
\def\zg{Z_{\rm g}}
\def\zm{Z_{\rm m}}
\def\zgm{Z_{\rm g,m}}
\def\zphi{Z_{\phi}}

\def\xg{X_{\rm g}}
\def\xm{X_{\rm m}}

\def\gtilde{\tilde{g}_0}
\def\mtilde{\widetilde{m}_0}

\def\msbar{{\rm \overline{MS\kern-0.05em}\kern0.05em}}
\def\smallmsbar{\overline{\hbox{\sixrm MS\kern-0.10em}}
                \hbox{\sixrm\kern0.10em}}
\def\lat{{\rm lat}}
\def\lmax{L_{\rm max}}

% PCAC and quark masses

\def\mup{m_{\rm u}}
\def\mdn{m_{\rm d}}
\def\ms{m_{\rm s}}
\def\Fpi{F_{\pi}}
\def\Gpi{G_{\pi}}
\def\mpi{m_{\pi}}
\def\FK{F_{\hbox{\sixrm K}}}
\def\fK{f_{\hbox{\sixrm K}}}
\def\fps{f_{\kern-1pt\hbox{\sixrm PS}}}
\def\fpi{f_{\pi}}
\def\fv{f_{\kern-1pt\hbox{\sixrm V}}}
\def\GK{G_{\hbox{\sixrm K}}}
\def\mK{m_{\hbox{\sixrm K}}}
\def\mKplus{m_{\kern-1pt\hbox{\sixrm K}^{+}}}
\def\mKnull{m_{\kern-1pt\hbox{\sixrm K}^{0}}}
\def\mKstar{m_{\hbox{\sixrm K}^{\ast}}}
\def\mps{m_{\kern-1pt\hbox{\sixrm PS}}}
\def\mv{m_{\hbox{\sixrm V}}}

\begin{titlepage}

\vbox{\vspace{1.0cm}}
\rightline{DESY 98-017}
\rightline{\thismonth\hspace{3pt}\thisyear}
\vspace{3.0cm}
\begin{center}
\large\bf
ADVANCED LATTICE QCD\hspace{1pt}%
\footnote[1]{Lectures given at the Les Houches Summer School
``Probing the Standard Model of Particle Inter\-actions",
July 28 -- September 5, 1997}
\end{center}
\vspace{1.0cm}

\begin{center}
\large
Martin L\"uscher 
\end{center}
\vspace{0.0cm}
\begin{center}
\it
Deutsches Elektronen-Synchrotron DESY \\
Notkestrasse 85, D-22603 Hamburg, Germany \\
E-mail: luscher@mail.desy.de
\end{center}

\end{titlepage}

\pagenumbering{roman}
\tableofcontents
\vfill\eject
\pagenumbering{arabic}

\section{Introduction}

At first sight quantum chromodynamics (QCD) 
appears to be a relatively simple theory 
with extensive symmetry properties and only few parameters. 
To the novice it must come as a surprise that the whole range of 
strong interaction phenomena can be described 
by such a beautiful mathematical structure.
Of course we do not know for sure whether this is really the case,
but the accumulated evidence strongly suggests this to be so.

The masses of the light hadrons and many other properties of these
particles are very precisely known from experiment. 
On the theoretical side, however, it is difficult to compute
these quantities from first principles, because of the non-linearities of 
the basic equations and because the coupling constant
is not small at low energies. Precision tests of 
QCD in this regime have consequently been rare and 
are essentially limited to those aspects of the theory that 
are determined by the (approximate) chiral and flavour symmetries.
It is evidently of great importance to overcome
this deficit and to verify that the same lagrangian which describes
the interactions between quarks and gluons at high energies also
explains the spectrum of light hadrons and their properties.

\subsection{Lattice QCD}

To be able to address this problem one first of all needs
a formulation of the theory which is mathematically well-defined 
at the non-perturbative level.
Such a framework is obtained by introducing a space-time lattice
and discretizing the fields and the action. The lattice
cuts off the high frequencies and makes the theory completely finite.
Eventually the continuum limit should be taken and one then 
encounters the usual ultra-violet divergencies which can be removed
through the usual parameter and operator renormalizations.
Lattice QCD may thus be regarded as a non-perturbative 
regularization of the theory with a momentum cutoff inversely proportional
to the lattice spacing. 

Today quantitative
results in lattice QCD are almost exclusively obtained using numerical
simulations. Such calculations proceed by choosing a finite lattice
which is sufficiently small that the quark and gluon fields can be
stored in the memory of a computer. Through a Monte Carlo algorithm
one then generates a representative ensemble of fields for the Feynman
path integral and extracts the physical quantities from ensemble
averages. Apart from statistical errors this method yields exact
results for the chosen lattice parameters and is hence suitable for
non-perturbative studies of QCD.

One should however not conclude that the theory can be solved
through numerical simulations alone.
Without physical insight and ingenious computational strategies
it would be quite impossible to obtain accurate 
results that can be compared with experiment.

\subsection{Continuum limit}

One of the most important and difficult issues in lattice QCD
is the continuum limit.
In practice the limit is taken by computing 
the quantities of interest for several values of the lattice spacing $a$ 
and extrapolating the results to $a=0$.
An obvious problem is then that 
one cannot afford to perform numerical simulations at 
arbitrarily small lattice spacings.
In hadron mass calculations, for example, current
lattice spacings are usually not much smaller than $0.1$ fm.
This will remain so for quite some time, because 
the simulation programs 
slow down proportionally to $a^5$ (or even a larger power of $a$)
if all other parameters are held fixed.

In view of these remarks, it is not totally obvious that 
the continuum limit can be reached in a reliable manner
from the presently accessible range of lattice spacings.
A detailed theoretical understanding of the approach to the limit is 
certainly required for this and  
extensive numerical studies are then needed to 
confirm (or disprove) the expected behaviour.

\subsection{Non-perturbative renormalization}

Non-perturbative renormalization is
another topic where one would not get very far
without theoretical input.
Traditionally renormalization is discussed in the framework of
perturbation theory, but when one is dealing with the
low-energy sector of QCD the renormalization constants
should be computed non-perturbatively.
In more physical terms, the question is
how precisely the perturbative regime of the theory 
is connected with the hadrons and their interactions. 
Evidently this is a fundamental aspect of QCD and not
particularly a lattice issue.

Non-perturbative renormalization involves large
scale differences and is hence not easily approached using 
numerical simulations, where the range of
physical distances covered by any single lattice is rather limited.
The problem has attracted a lot of attention recently and
various ways to solve it have been proposed.

\subsection{Contents \& aims of the lectures}

The continuum limit and non-perturbative renormalization are the 
main themes of the lectures. In both areas significant progress has
been made in the past few years.
Chiral symmetry plays an important r\^ole in some of these developments
and the way in which this symmetry is realized on the lattice is 
hence also discussed.

Another item which is given considerable weight are finite-size
techniques. Such methods are probably unavoidable if one would like
to overcome the limitations imposed by the currently accessible
lattices. Some of them have already proved to be very powerful
and the aim of the lectures partly is to 
illustrate this and to provide a better understanding of the underlying
physical picture. 

The lectures are self-contained, but some familiarity with the basic
concepts in lattice gauge theory is assumed throughout.
It goes without saying that many interesting theoretical 
developments in lattice QCD are not covered and that
the selected topics merely reflect the knowledge and preferences of the 
lecturer.

\subsection*{Acknowledgements}

I would like to thank the organizers of the school, particularly
Rajan Gupta, for the opportunity to give these lectures 
and for some very pleasant and stimulating weeks at Les Houches.
Much of the material presented here is based on work done
by the ALPHA collaboration. 
I wish to thank the members of the collaboration
for their long-term commitment and an enjoyable team-work.

\section{Lattice effects and the continuum limit}

Nearly 20 years ago Symanzik set out to study 
the nature of the continuum limit
in perturbation theory~[\Cite{SymanzikI}]. He showed that 
lattice theories can be 
described through an effective continuum theory 
in which the lattice spacing dependence is made explicit.
This work later led him to propose a method to accelerate
the approach to the continuum limit, which is now known as
the Symanzik improvement programme~[\Cite{SymanzikII},\Cite{SymanzikIII}].
In this section we discuss the effective continuum theory for lattice QCD.
A particular form of improvement will be considered in section~3.

\subsection{Preliminaries}

In the standard formulation of lattice QCD (which goes back
to Wilson's famous paper of 1974~[\Cite{Wilson}])
the quark fields $\psi(x)$ reside on the sites $x$ of the lattice
and carry colour, flavour and Dirac indices 
as in the continuum theory. The gauge field is represented
through a field of SU(3) matrices $U(x,\mu)$ where $\mu=0,\ldots,3$
labels the space-time directions. None of these is special since
we are in euclidean space, but $\mu=0$ is conventionally associated
with the time direction.

Under a gauge transformation $\Lambda(x)$ the fields transform
according to 
\begin{displaymath}
  \psi(x)\to\Lambda(x)\psi(x),
  \qquad
  U(x,\mu)\to\Lambda(x)U(x,\mu)\Lambda(x+a\hat{\mu})^{-1},
\end{displaymath}
where $\hat{\mu}$ denotes the unit 
vector in direction $\mu$.
The forward difference operator
\begin{displaymath}
  \nab{\mu}\psi(x)={1\over a}\bigl[
  U(x,\mu)\psi(x+a\hat{\mu})-\psi(x)
  \bigr]
\end{displaymath}
is hence gauge covariant. In the same way 
a backward difference operator $\nabstar{\mu}$ may be
defined and the lattice Dirac operator is then given by
\begin{equation}
  D=\frac{1}{2}\left\{
  \dirac{\mu}(\nabstar{\mu}+\nab{\mu})
  -a\nabstar{\mu}\nab{\mu}\right\}.
  \label{diracoperator}
\end{equation}
The second term (which is referred to as the Wilson term) 
is included to avoid the well-known ``doubler" problem.
It assigns a mass proportional to $1/a$ 
to all modes with momenta of this order
while the low-momentum modes are affected by
corrections vanishing proportionally to $a$.
In particular, only these modes 
survive in the continuum limit.

The QCD action on the lattice may now be written in the form
\begin{equation}
  S={1\over g_0^2}\sum_{p}\tr\{1-U(p)\}
  +a^4\sum_{x}\psibar(x)(D+m_0)\psi(x)
  \label{latticeaction}
\end{equation}
with $g_0$ being the bare gauge coupling and $m_0$
the bare quark mass. For simplicity we shall assume
that the quark mass is the same for all flavours.
The sum in the gauge field part of the action runs over all oriented
plaquettes~$p$ and $U(p)$ denotes the product of the 
gauge field variables around~$p$.

\exercise{List all continuous and discrete 
symmetries of the lattice action
and write down the corresponding transformations of the fields}

\subsection{Lattice effects in perturbation theory}

In lattice QCD one is primarily interested in the non-perturbative
aspects of the theory. 
Perturbation theory can, however, give important structural insights
and it has proved useful to study the approach to the continuum limit
in this framework. A remarkable result in this connection is
that the existence of the limit has been rigorously established
to all orders of the expansion~[\Cite{Reisz},\Cite{LesHouchesI}]. 

The Feynman rules on the lattice are derived straightforwardly
from the lattice action by introducing the gluon field $A_{\mu}(x)$ through
\begin{displaymath}
  U(x,\mu)=\exp\left\{g_0aA_{\mu}(x)\right\},
  \qquad
  A_{\mu}(x)=A_{\mu}^a(x)\lambda^a,
\end{displaymath}
and expanding in powers of $g_0$.
One also needs to fix the gauge but this will not be discussed here
(details can be found in  ref.~[\Cite{LesHouchesI}] for example). 
Compared to the usual Feynman rules, an important difference is
that the propagators and vertices are relatively complicated functions
of the momenta and of the lattice spacing $a$.
In particular, at tree-level all the lattice spacing dependence
arises in this way.

A simple example illustrating this is the quark-gluon vertex

\rightline{
\begin{minipage}{6cm}
\epsfxsize=1.5cm\vspace{0.4cm}\hspace{4.5cm}\epsfbox{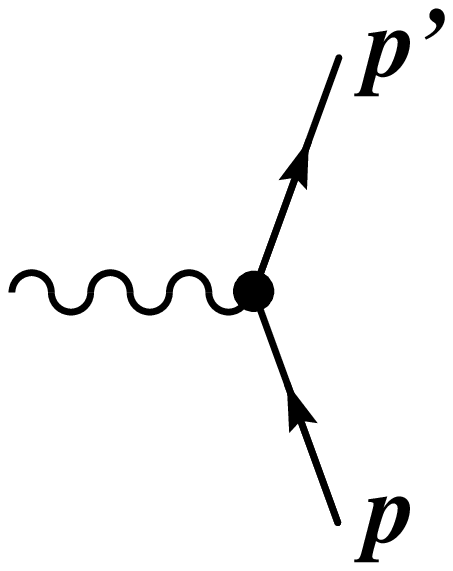}
\end{minipage}
\begin{minipage}{10cm}
\begin{equation}
  =\;-g_0\lambda^a\left\{
   \dirac{\mu}-\frac{i}{2}a(p+p')_{\mu}+\rmO(a^2)\right\}.
   \hspace{2.5cm}
   \label{qgvertex}
\end{equation}
\end{minipage}}

\vspace{0.5cm plus 0.05cm}
\noindent
It is immediately clear from this expression that
the leading lattice corrections to the continuum term
can be quite large even if the quark momenta $p$ and $p'$ are 
well below the momentum cutoff $\pi/a$. 
Moreover the corrections violate chiral symmetry,
a fact that has long been a source of concern
since many properties of low-energy QCD depend on this symmetry.

Lattice Feynman diagrams with $l$ loops
and engineering dimension $\omega$ can be expanded in 
an asymptotic series 
of the form~[\Cite{SymanzikI},\Cite{SymanzikII}]
\begin{displaymath}
  a^{-\omega}\sum_{k=0}^{\infty}\sum_{j=0}^l c_{kj}a^k\left[\ln(a)\right]^j.
\end{displaymath}
After renormalization the negative powers in the lattice spacing
and the logarithmically divergent terms cancel in the sum of all diagrams.
The leading lattice corrections thus
vanish proportionally to $a$ (up to logarithms)
at any order of perturbation theory.

\exercise{Work out the exact expression for the quark-gluon vertex and 
verify eq.~(\ref{qgvertex})}

\begin{figure}[t]
\vspace{-2.3cm}
\centerline{\epsfig{file=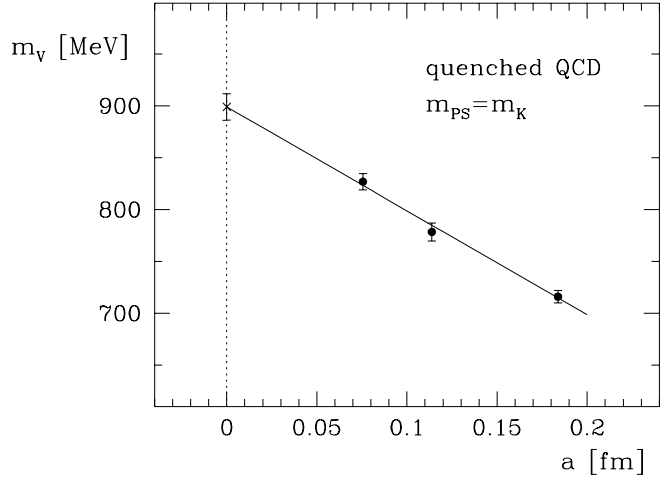,width=11.0cm}}
\vspace{-1.2cm}
\caption{Calculated values of 
the vector meson mass (full circles) and
linear extrapolation to the continuum limit (cross).
Simulation data from Butler et al.~(GF11 collab.)~[\Cite{WeingartenI}].}
\end{figure}

\subsection{Non-perturbative test for lattice effects}

Sizeable lattice effects are also observed 
at the non-perturbative level when calculating hadron masses, for example.
An impressive demonstration of this is obtained as follows.
Let us suppose that the quark mass is adjusted so that
the mass of the pseudo-scalar mesons coincides with
the physical kaon mass. This sets the quark mass to about half
the strange quark mass and one thus expects
that the mass of the lightest vector meson is given by
\begin{displaymath}
  \mv\simeq\mKstar=892\,\MeV.
\end{displaymath}
Computations of the meson masses using numerical simulations
however show that this is not the case at 
the accessible lattice spacings (see fig.~1).
Instead one observes a strong dependence on the lattice spacing
and it is only after extrapolating the data to $a=0$ that one ends up
with a value close to expectations.

The numerical simulations quoted here have been performed in
``quenched QCD" where sea quark effects are neglected.
It seems unlikely, however, that a very different result would
be obtained in the full theory. 
Perturbation theory and the calculations in quenched QCD rather suggest
that large lattice effects already arise from the 
interaction between the gluon field and the valence quarks
and these are not suppressed when the sea quarks are included.

\subsection{Effective continuum theory}

In phenomenology it is well-known that the effects of as yet
undetected substructures or heavy particles may be described by adding
higher-dimensional interaction terms to the Stan\-dard Model lagrangian.
From the point of view of an underlying more complete theory, the
Standard Model together with the added terms is then a low-energy
effective theory. 

A similar situation occurs in lattice QCD, where
the momentum cutoff may be regarded (in a purely mathematical sense)
as a scale of new physics. The associated low-energy effective theory
is a continuum theory with action~[\Cite{SymanzikII},\Cite{SymanzikIII}]
\begin{equation}
  \Seff=\int\rmd^4x \left\{
  {\cal L}_0(x)+a{\cal L}_1(x)+a^2{\cal L}_2(x)+\ldots\right\},
  \label{effaction}
\end{equation}
where ${\cal L}_0$ denotes the continuum QCD lagrangian
and the ${\cal L}_k$'s, $k\geq1$, are linear combinations of 
local operators of dimension $4+k$.
The dimension counting here includes the (non-negative) powers of the
quark mass $m$ by which some of the fields may be multiplied.
From the list of all possible such fields
only those need to be considered which are invariant under
the symmetries of the lattice theory. 
Many terms can also be eliminated using partial integration.

In the following we shall be mainly interested in 
the lattice corrections of order~$a$.
All other corrections are proportional to higher powers of~$a$
and are thus expected to be less important at
small lattice spacings.
Taking the symmetries of the lattice action
into account, it is not difficult to show that 
the effective lagrangian ${\cal L}_1$
may be written as a linear combination of the fields
\begin{eqnarray*}
  \op{1}&=&\psibar\,i\sigma_{\mu\nu}F_{\mu\nu}\psi,
  \label{op1}
  \\
  \noalign{\vskip1ex}
  \op{2}&=&\psibar\,D_{\mu}D_{\mu}\psi
          +\psibar\,\lvec{D}_{\mu}\lvec{D}_{\mu}\psi,
  \label{op2}
  \\
  \noalign{\vskip1ex}
  \op{3}&=&m\tr\!\left\{F_{\mu\nu}F_{\mu\nu}\right\},
  \label{op3}
  \\
  \noalign{\vskip1ex}
  \op{4}&=&m\left\{\psibar\,\dirac{\mu}D_{\mu}\psi
  -\psibar\,\lvec{D}_{\mu}\dirac{\mu}\psi\right\},
  \label{op4}
  \\
  \noalign{\vskip1ex}
  \op{5}&=&m^2\,\psibar\psi,
  \label{op5}
\end{eqnarray*}
where $F_{\mu\nu}$ denotes the gauge field strength and $D_{\mu}$
the gauge covariant partial derivative.

In the effective continuum theory, renormalized lattice fields
are represented through effective fields of the form
\begin{equation}
  \phieff=\phi_0+a\phi_1+a^2\phi_2+\ldots
  \label{efffield}
\end{equation}
The fields $\phi_0$, $\phi_1,\ldots$ which can occur here
must have the appropriate dimension and transform
under the lattice symmetries in the same way 
as the lattice field which is being represented.
As an example let us consider 
the isovector axial current
\begin{equation}
  A^a_{\mu}(x)=\psibar(x)\dirac{\mu}\dirac{5}\frac{1}{2}\tau^a\psi(x).
  \label{axialcurrent}
\end{equation}
In this case the leading correction $\phi_1$
can be shown to be a linear combination of the fields
\begin{eqnarray*}
  (\op{6})_{\mu}^a&=&
  \psibar\,\dirac{5}\frac{1}{2}\tau^a\sigma_{\mu\nu}D_{\nu}\psi-
  \psibar\,\lvec{D}_{\nu}\sigma_{\mu\nu}\dirac{5}\frac{1}{2}\tau^a\psi,
  \\
  \noalign{\vskip1ex}
  (\op{7})_{\mu}^a&=&
  \partial_{\mu}\left\{\psibar\,\dirac{5}\frac{1}{2}\tau^a\psi\right\},
  \\
  \noalign{\vskip1ex}
  (\op{8})_{\mu}^a&=&
  m\psibar\,\dirac{\mu}\dirac{5}\frac{1}{2}\tau^a\psi.
\end{eqnarray*}
For other low-dimensional fields, such as the vector currents and 
the scalar and pseudo-scalar densities, the general form of the 
O($a$) corrections is also easily found. The number of terms
is however rapidly increasing when one considers operators
involving four quark fields or higher derivatives.

\exercise{Prove that any local field of dimension~5
respecting the symmetries of the lattice theory can be
written as a linear combination of the basis $\op{1},\ldots,\op{5}$ up
to derivative terms}

\subsection{Scope of the effective continuum theory}

So far we did not specify how precisely the lattice theory
is related to the effective continuum theory.
One must be a bit careful here, because additional lattice effects
can arise when integrating correlation functions
over short-distance regions. Such effects are characteristic
of the correlation functions considered and are not 
accounted for by the effective action and the effective fields.
In the following attention will thus be restricted 
to position space correlation functions 
at non-zero (physical) distances.
This restriction is not very severe, since
most physical quantities, including the hadron masses and matrix
elements of local operators between particle states, 
can be extracted from such correlation functions.

So let us consider some
local gauge invariant field $\phi(x)$ constructed 
from the quark and gauge fields on the lattice.
For simplicity 
we assume that $\phi(x)$
does not mix with other fields under renormalization.
We then expect that the connected renormalized
$n$-point correlation functions 
\begin{displaymath}
  G_n(x_1,\ldots,x_n)=(\zphi)^n
  \left\langle\phi(x_1)\ldots\phi(x_n)\right\rangle_{\rm con}
\end{displaymath}
have a well-defined continuum limit, provided the renormalization factor
$\zphi$ is chosen appropriately and if all points $x_1,\ldots,x_n$ are 
kept at non-zero distances from one another.

In the effective continuum theory, the lattice correlation functions 
are represented through an asymptotic expansion
\begin{eqnarray}
  G_n(x_1,\ldots,x_n)&=&
  \left\langle\phi_0(x_1)\ldots\phi_0(x_n)\right\rangle_{\rm con}
  \nonumber
  \\
  \noalign{\vskip1.5ex}
  &&
  -a\int\rmd^4y\,\left\langle\phi_0(x_1)\ldots\phi_0(x_n)
  {\cal L}_1(y)\right\rangle_{\rm con}
  \nonumber
  \\
  \noalign{\vskip0.8ex}
  &&
  +a\sum_{k=1}^n
  \left\langle\phi_0(x_1)\ldots\phi_1(x_k)\ldots\phi_0(x_n)
  \right\rangle_{\rm con}+\rmO(a^2),
  \label{cfunction}
\end{eqnarray}
where the expectation values on the right-hand side are to be taken
in the continuum theory with lagrangian ${\cal L}_0$.
The second term is the contribution of the O($a$)
correction in the effective action. Note that 
the integral over $y$ in general diverges 
at the points $y=x_k$. 
A subtraction prescription must 
hence be supplied. The precise way in which this happens is unimportant,
because the arbitrariness that one has amounts to a local operator
insertion at these points, i.e.~to a redefinition of the field $\phi_1(x)$.

It should be emphasized at this point 
that not all the dependence on the lattice
spacing comes from the explicit factors of $a$ in 
eq.~(\ref{cfunction}).
The other source of $a$-dependence are 
the operators $\phi_1$ and ${\cal L}_k$, which are linear
combinations of some basis of fields. While the basis elements are 
independent of $a$, the coefficients need not be so, although
they are expected to vary only slowly with $a$. 
In perturbation theory the coefficients are calculable polynomials 
in $\ln(a)$.

\subsection{Elimination of redundant terms}

The list of operators given above for the O($a$) lattice corrections
can be reduced using the field equations.
Formal application of the field equations yields the relations
\begin{equation}
  \op{1}-\op{2}+2\op{5}=0,
  \qquad
  \op{4}+2\op{5}=0,
  \label{oprelations}
\end{equation}
which allow one to eliminate $\op{2}$ and $\op{4}$, for example.
$\op{6}$ is also redundant since it can be expressed as a linear 
combination of $\op{7}$ and $\op{8}$. 

There are two technical remarks which should be made here.
The first is that in euclidean correlation functions
the field equations are only valid up to contact terms.
In eq.~(\ref{cfunction}) these can arise when $y$ gets close to 
one of the points $x_k$. The possible contact terms there however
amount to an operator insertion, which can be compensated
by a redefinition of $\phi_1$. The latter thus depends
on exactly which basis of operators has been selected for the 
O($a$) effective action. Note that when the field equations
are used to simplify $\phi_1$, no contact terms can arise since
all points $x_k$ keep away from each other.

The second observation is that renormalization and
operator mixing must be taken into account when deriving
operator relations from the field equations. 
In particular, the naive relations, eq.~(\ref{oprelations}),
are only valid at tree-level of perturbation theory.
At higher orders they
are replaced by linear combinations of all basis elements 
with coefficients that depend on the coupling and the 
chosen normalization conditions.
Barring singular events, it is then still possible to eliminate
$\op{2}$, $\op{4}$ and $\op{6}$.

\exercise{Work out the free quark propagator on the lattice
and check that its dependence on the lattice spacing 
matches with what is expected
from the effective theory}

\subsection{Concluding remarks}

Through the effective continuum theory
a better understanding of the approach to the continuum limit is achieved.
In particular, omitting terms that are redundant
or which amount to renormalizations of the coupling and the quark mass,
the leading lattice correction in the effective action is 
proportional to the Pauli term $\psibar\,i\sigma_{\mu\nu}F_{\mu\nu}\psi$.
The lattice thus assigns an anomalous
colour-magnetic moment of order $a$ to the quarks.
Very many more terms contribute to ${\cal L}_2$ 
and a simple physical interpretation is not easily given.
The pattern of the lattice effects of order $a^2$ 
should hence be expected to be rather complicated.

\section{O($a$) improvement}

While the general form
of the effective continuum theory 
is determined by the field content and the symmetries of 
the lattice theory,
the coefficients multiplying the 
basis fields in the effective action and the effective fields 
depend on how precisely the latter has been set up.
With an improved lattice action and improved expressions for 
the local composite fields we may hence be able 
to reduce the size of the lattice effects and thus to accelerate
the convergence to the continuum limit
[\Cite{SymanzikII},\Cite{SymanzikIII}].
The idea is similar to using higher-order discretization schemes
for the numerical solution of partial differential equations, but
the situation is not quite as simple because of the non-linearities
and the associated renormalization effects.

Improvement can be implemented in different ways 
and there is currently no single preferred way to proceed.
The subject has been reviewed by Niedermayer~[\Cite{ImpReview}]
and many interesting contributions have been made since then.
O($a$) improvement is a relatively modest approach, where one attempts
to cancel all lattice effects of order~$a$.
This development started a long time ago [\Cite{OnShell}--\Cite{Wohlert}]
and the method is now becoming increasingly usable 
although there are still a few loose ends.

\subsection{O($a$) improved action}

In the following our aim
is to construct an improved lattice action 
by adding a suitable
counterterm to the Wilson action eq.~(\ref{latticeaction}).
The counterterm should be chosen such that 
the order $a$ term in the action of the effective continuum
theory is cancelled.
We are only interested in position space correlation functions 
of the type discussed above and may hence
assume that the effective lagrangian ${\cal L}_1$
is a linear combination of the reduced basis of fields 
$\op{1}$, $\op{3}$ and $\op{5}$. 
It is then quite obvious that ${\cal L}_1$
can be made to vanish by 
adding a counterterm of the form
\begin{displaymath}
  \delta S=a^5\smash{\sum_x}\left\{
  c_1\ophat{1}(x)+c_3\ophat{3}(x)+c_5\ophat{5}(x)
  \right\},
\end{displaymath}
where $\widehat{\cal O}_k$ is 
some lattice representation of the field ${\cal O}_k$.

Apart from renormalizations of the bare parameters 
and adjustments of the coefficients $c_k$, 
the discretization ambiguities that one has here are of order $a^2$.
In particular, we may choose to represent 
the fields $\tr\!\left\{F_{\mu\nu}F_{\mu\nu}\right\}$ and $\psibar\psi$
by the pla\-quette field and the local scalar density that already
appear in the Wilson action.
The terms proportional $\ophat{3}$ and $\ophat{5}$
then amount to a rescaling of the bare coupling and mass
by factors of the form $1+\rmO(am)$. 
We ignore this rescaling for the time being and
return to the issue later when we discuss the renormalization of 
the improved theory.

\begin{figure}[t]
\vspace{0.0cm}
\centerline{\epsfig{file=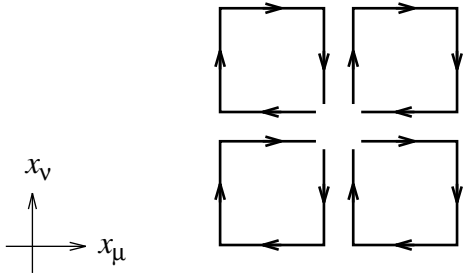,width=6.0cm}}
\vspace{0.0cm}
\caption{
Graphical representation of the products of gauge field
variables contributing to the lattice field tensor 
[eq.~(\ref{fieldtensor})]. The point $x$
is at the centre of the diagram where all loops start and end.}
\end{figure}

For the improved action we thus obtain
\begin{equation}
  S=S_{\rm Wilson}+
  a^5\smash{\sum_{x}}\,
  \csw\,
  \psibar(x)\frac{i}{4}\sigma_{\mu\nu}\widehat{F}_{\mu\nu}(x)\psi(x),
  \label{improvedaction}
\end{equation}
where $\widehat{F}_{\mu\nu}$ is a lattice representation of the 
gluon field tensor. 
Recall that the product of the gauge field
variables around a plaquette in the $(\mu,\nu)$--plane is equal to 
$1+a^2F_{\mu\nu}+\ldots$ in the classical continuum limit.
A symmetric definition of 
the lattice field tensor is hence given by
\begin{equation}
  \widehat{F}_{\mu\nu}(x)={1\over8a^2}\left\{
  Q_{\mu\nu}(x)-Q_{\nu\mu}(x)\right\}
  \label{fieldtensor}
\end{equation}
with $Q_{\mu\nu}(x)$ being the sum of 
the plaquette loops shown in fig.~2.

The improved action, eq.~(\ref{improvedaction}), first appeared
in a paper by Sheikholeslami and Wohlert in 1985 [\Cite{SW}].
It did not receive too much attention at the time, because
systematic studies of lattice effects were not feasible with the available
computers. The situation has now changed and there is general agreement
that improvement is useful or even necessary,
particularly in full QCD where simulations
are exceedingly expensive in terms of computer time.

To achieve the desired improvement,
the coefficient $\csw$ multiplying the O($a$) counterterm in 
the improved action should be chosen appropriately.
For various technical reasons this is a non-trivial task
and the tuning of $\csw$ has long been considered 
a weak point of the method.
Sheikholeslami and Wohlert calculated the coefficient at tree-level of
perturbation theory and Wohlert [\Cite{Wohlert}]
later obtained the one-loop formula
\begin{equation}
  \csw=1+0.2659\times g_0^2+\rmO(g_0^4).
  \label{cswoneloop}
\end{equation}
More recently a strategy has been developed which 
allows one to compute $\csw$ non-perturbatively using numerical 
simulations [\Cite{paperI}].
Chiral symmetry plays an important r\^ole in this context
and the further discussion of this issue is hence postponed
to section~5.

\exercise{Write down the expression for the lattice field tensor
explicitly and check that it transforms under the lattice symmetries
in the expected way}

\subsection{Improvement of local fields}

Through the improvement of the action one achieves that
on-shell quantities
such as particle masses and scattering
matrix elements approach the continuum limit 
with a rate proportional to $a^2$.
As can be seen from eq.~(\ref{cfunction}),
some terms of order $a$ however
remain uncancelled in correlation functions
of local fields. 
For a complete improvement of such correlation functions
one also needs to improve the fields and to specify appropriate 
renormalization conditions.

Improved fields are constructed by adding 
local counterterms to the unimproved fields in such a way that the
order $a$ term in the associated effective fields is cancelled.
The counterterms are linear combinations of a basis of operators
with the appropriate dimension and symmetry properties.
In the case of the isovector axial current, for example,
the basis consists of a lattice representation of the fields 
$\op{6}$, $\op{7}$ and $\op{8}$.
The first of these is redundant and may be dropped
without loss. The term associated with $\op{8}$
amounts to a renormalization of the current 
and is hence only meaningful when a renormalization condition
has been imposed. Since we have not done so at this point,
we drop this term too and shall address the issue again 
in the next subsection.

We are then left with a single term and 
the improved axial current becomes
\begin{equation}
  (\aimpr)^a_{\mu}=A^a_{\mu}+\ca a\drvtilde{\mu}P^a,
  \label{improvedaxialcurrent}
\end{equation}
where $\drvtilde{\mu}$ denotes
the average of the forward and backward difference operators and  
\begin{equation}
  P^a(x)=\psibar(x)\dirac{5}\frac{1}{2}\tau^a\psi(x)
  \label{axialdensity}
\end{equation}
the isovector pseudo-scalar density. 
The coefficient $\ca$ depends on $g_0$
and should be chosen so as 
to achieve the O($a$) improvement of the correlation 
functions of the axial current. In perturbation theory
one obtains [\Cite{paperII}]
\begin{displaymath}
  \ca=-0.00756\times g_0^2+\rmO(g_0^4)
\end{displaymath}
and in quenched QCD 
the coefficient is now also known non-perturbatively
[\Cite{paperIII}].

\exercise{Show that the pseudo-scalar density, 
eq.~(\ref{axialdensity}), is already improved,
i.e.~no O($a$) counterterm is needed in this case}

\subsection{Renormalization}

When deriving the improved action and the expression 
eq.~(\ref{improvedaxialcurrent}) for the improved axial current,
all terms that amount to renormalizations of the coupling constant,
the quark mass and the current by factors of the form $1+\rmO(am)$
have been dropped. The reason for this has been that 
such factors should better be discussed in the context of renormalization
which we now do.

In QCD with only light quarks it is technically advantageous
to employ so-called mass-independent renormalization schemes.
If we ignore O($a$) improvement for while, the renormalized
coupling and quark mass in such schemes
are related to the bare parameters through
\begin{eqnarray*}
  \gr^2&=&
  g_0^2\zg(g_0^2,a\mu),
  \\
  \noalign{\vskip1.0ex}
  \mr&=&
  \mq\zm(g_0^2,a\mu), 
  \qquad
  \mq=m_0-\mc.
\end{eqnarray*}
The important point here is that the renormalization factors
only depend on the normalization mass $\mu$
and the coupling, but not on the quark mass.
Note that an
additive quark mass renormalization $\mc$ (equal to $1/a$ times
some function of $g_0$)
is required on the lattice, because chiral symmetry
is not preserved.

In the improved theory the definition of the renormalized parameters
must be slightly modified, as otherwise
one may end up with uncancelled cutoff effects
of order $a\mq$ in some correlation functions. 
From our discussion above one infers that we should
first rescale the bare parameters according to
\begin{eqnarray*}
  \gtilde^2&=&
  g_0^2\left(1+\bg a\mq\right),
  \label{imprg}
  \\
  \noalign{\vskip1.0ex}
  \mqtilde&=&
  \mq\left(1+\bm a\mq\right),
  \label{imprm}
\end{eqnarray*}
and then define the renormalized coupling and mass through
\begin{eqnarray*}
  \gr^2&=&
  \gtilde^2\zg(\gtilde^2,a\mu),
  \label{imprschemeg}
  \\
  \noalign{\vskip1.0ex}
  \mr&=&
  \mqtilde\zm(\gtilde^2,a\mu).
  \label{imprschemem}
\end{eqnarray*}
The coefficients $\bg$ and $\bm$ depend on $g_0$ and must be chosen
appropriately to cancel any remaining lattice effects of order $a\mq$.
They have been worked out to one-loop order of perturbation theory
[\Cite{bgOneloop},\Cite{bmOneloop}]
and $\bm$ is now also known non-perturbatively in quenched
QCD [\Cite{bmNonpert}].

Factors of the form $1+\rmO(a\mq)$ must
also be included in the definition of renormalized improved fields. The 
renormalized axial current, for example, is given by
\begin{displaymath}
  (\ar)^a_{\mu}=
  \za(1+\ba a\mq)(\aimpr)^a_{\mu}
\end{displaymath}
and similar formulae apply to any multiplicatively renormalizable field.
Even though the renormalized fields are written in this way, it
should be noted that the $b$-coefficients do not depend on the 
renormalization conditions which are being imposed. The latter
should be set up at zero quark masses and only fix the $Z$-factors.

\exercise{Consider the free quark theory and prove that 
$\bm=-\frac{1}{2}$ in this case. Determine the renormalization factor needed
to improve the position space correlation functions of the quark field}

\subsection{Is O($a$) improvement effective?}

Numerical studies of lattice QCD
using the improved action eq.~(\ref{improvedaction})
have been initiated a few years ago
[\Cite{HeatlieEtAl}--\Cite{UKQCDI}].
The additional computational work
required for the improvement term 
generally takes less than 20\% of the total computer time
spent for the simulations, also in the unquenched case [\Cite{KarlChuan}].
An important question which one would like to answer with 
these studies is whether the lattice effects
on the quantities of physical interest 
are indeed significantly reduced at the accessible lattice spacings.
For the non-perturbatively determined values of $\csw$
several collaborations
have recently set out to check this~[\Cite{QCDSFa}--\Cite{SCRIb}],
but it is too early to draw definite conclusions.

For illustration let us again consider
the calculation of the vector meson mass $\mv$ discussed in 
subsection~2.3. 
A preliminary analysis of simulation results from
the UKQCD collaboration gives, for the O($a$) improved theory,
$\mv=924(17)$ MeV at $a=0.098$ fm and $\mv=932(26)$ MeV 
at $a=0.072$ fm [\Cite{UKQCDb},\Cite{WittigI}].
These numbers do not show any significant 
dependence on the lattice spacing and they are also compatible with 
the value $\mv=899(13)$ that one obtains through extrapolation
to $a=0$ of the results from the unimproved theory (left-most point in
fig.~1). Data at larger values of the lattice spacing recently
published by Edwards et al.~[\Cite{SCRIb}] corroborate these findings.

Other quantities that are being studied include the pseudo-scalar
and vector meson decay constants and the renormalized quark masses.
Compared to the hadron mass calculations, the situation in these
cases is significantly more complicated, because one also needs to
determine the appropriate renormalization factors and 
the improved expressions for the isovector axial and vector currents. 
The experience accumulated so far
suggests that the residual lattice effects are indeed
small if $a\leq0.1$ fm.
Further confirmation is however still needed.

\subsection{Synthesis}

At sufficiently small lattice spacings 
the effective continuum theory provides an elegant description of 
the approach to the continuum limit.
Whether the currently accessible lattice spacings
are in the range where the effective theory applies
is not immediately clear,
but the observed pattern of the lattice spacing dependence in the 
unimproved theory and the fact that O($a$) improvement 
appears to work out strongly indicate this to be so.
Very much smaller lattice spacings are then not required
to reliably reach the continuum limit. 
It is evidently of great importance to put this conclusion
on firmer grounds by continuing and extending the ongoing 
studies of O($a$) improvement and other forms of improvement.

\vfill\eject

\section{Chiral symmetry in lattice QCD}

Many properties of low-energy QCD can be explained through
the flavour and chiral symmetries of the theory.
For the analysis of the weak interactions of the quarks,
these symmetries are also very important and 
the fact that chiral symmetry is not exactly preserved 
on the lattice thus is a source of concern.
In this section we discuss how precisely the symmetry is violated
and indicate how to make use of it even though it is only approximate.

\subsection{Chiral Ward identities}

We first consider the theory in the continuum limit and proceed
formally, i.e.~without paying attention to the proper definition
of the correlation functions that occur.
We assume that there is an isospin doublet of
quarks with equal mass $m$ and 
study the associated chiral transformations.

In euclidean space the expectation value of any product $\cal O$
of local composite fields is given by the functional integral
\begin{displaymath}
  \langle{\cal O}\rangle={1\over{\cal Z}}
  \int_{\rm fields}{\cal O}\,\rme^{-S}.
\end{displaymath}
Starting from this expression the general chiral Ward identity
is obtained by applying an 
infinitesimal chiral rotation
\begin{displaymath}
  \delta\psi(x)=\omega^a(x)\frac{1}{2}\tau^a\dirac{5}\psi(x),
  \qquad
  \delta\psibar(x)=\omega^a(x)\psibar(x)\dirac{5}\frac{1}{2}\tau^a,
\end{displaymath}
to the integration variables.
In these equations $\tau^a$ denotes a Pauli matrix acting on the 
isospin indices of the quark field and $\omega^a(x)$ is a
smooth function which vanishes outside some bounded region $R$.
Since the Pauli matrices are traceless, it is obvious that
the integration measure is invariant under the transformation
and we thus conclude that
\begin{equation}
  \langle\delta S\,{\cal O}\rangle=\langle\delta{\cal O}\rangle.
  \label{wardidentity}
\end{equation}
The chiral variation of the action and of the local fields of interest
are easily worked out. In particular, for the variation of the action 
one finds
\begin{equation}
  \delta S=\int\rmd^4x\,\omega^a
  \left\{-\partial_{\mu}A^a_{\mu}+2mP^a\right\},
  \label{deltaS}
\end{equation}
where $A^a_{\mu}$ and $P^a$ denote the isovector axial current and density
[eqs.~(\ref{axialcurrent}),(\ref{axialdensity})].
The derivation of this result involves a partial integration
and it has been important here that $\omega^a(x)$ smoothly goes to zero
outside some bounded domain
as otherwise a boundary term might arise.

By inserting 
particular products of fields ${\cal O}$ and by choosing
different functions $\omega^a(x)$,
many useful relations 
(collectively referred to as the chiral Ward identities) are thus obtained.
A few special cases will be considered in the following paragraphs.

\exercise{Show that the axial current $A^a_{\mu}$
and the isovector vector current 
$V^a_{\mu}=\psibar\dirac{\mu}\frac{1}{2}\tau^a\psi$ transform into each other
under infinitesimal chiral rotations}

\subsection{PCAC relation}

Let us first assume that 
the fields in the product $\cal O$
are localized outside the region $R$ as shown in fig.~3.
The variation $\delta{\cal O}$ vanishes under these conditions
and eq.~(\ref{wardidentity}) reduces to 
$\langle\delta S\,{\cal O}\rangle=0$. 
Since this equation holds for all functions $\omega^a(x)$ with support
in $R$, it follows that 
\begin{equation}
  \langle\partial_{\mu}A^a_{\mu}(x){\cal O}\rangle=
  2m\langle P^a(x){\cal O}\rangle.
  \label{PCACa}
\end{equation}
At zero quark mass this identity expresses
the conservation of the axial current and it is hence referred
to as the partially conserved axial current (PCAC) relation.

It should be emphasized that eq.~(\ref{PCACa}) holds irrespectively
of the product $\cal O$ of fields as long as 
these are localized in a region not containing $x$. Of course this
is just a reflection of the fact that the PCAC relation becomes an 
operator identity when the theory is set up in Minkowski space.
Different choices of $\cal O$ then correspond to 
considering different matrix elements of the operator relation.

\exercise{Derive the general Ward identity associated with the
$\SUtwo$ flavour group. What happens if the quarks have unequal masses?}

\begin{figure}[t]
\vspace{0.5cm}
\centerline{\epsfig{file=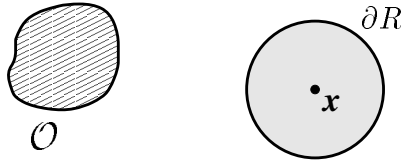,width=5.4cm}}
\vspace{0.5cm}
\caption{Choice of the region~$R$ when deriving the PCAC relation.
The product of fields~$\cal O$ is assumed to be
localized in the shaded area away from~$R$.}
\end{figure}

\subsection{Integrated Ward identities}

Another class of identities is obtained
when $\cal O$ factors into two products of fields, 
${\cal O}_{\rm int}$ and ${\cal O}_{\rm ext}$, localized
inside and outside the region~$R$ respectively (see fig.~4).
In the limit where $\omega^a(x)$ goes to a constant in $R$,
eq.~(\ref{wardidentity}) then becomes
\begin{eqnarray}
  &&\int_{\partial R}\rmd\sigma_{\mu}(x)\,\omega^a
  \langle 
  A^a_{\mu}(x) {\cal O}_{\rm int} {\cal O}_{\rm ext} 
  \rangle=
  \nonumber
  \\
  \noalign{\vskip1.5ex}
  &&
  \qquad\quad-\left\langle 
  \delta{\cal O}_{\rm int}\,{\cal O}_{\rm ext} 
  \right\rangle
  +2m\int_R\rmd^4x\,\omega^a
  \left\langle 
  P^a(x) {\cal O}_{\rm int} {\cal O}_{\rm ext} 
  \right\rangle,
  \label{intwardid}
\end{eqnarray}
where the integration on the left-hand side runs over the boundary of $R$.
So if we set $m=0$ and insert ${\cal O}_{\rm int}=A_{\nu}^b(y)$,
for example, a little algebra yields
\begin{equation}
  \int_{\partial R}\rmd\sigma_{\mu}(x)\, 
  \langle 
  A^a_{\mu}(x) A^b_{\nu}(y) {\cal O}_{\rm ext} 
  \rangle
  =
  i\epsilon^{abc}
  \left\langle 
  V^c_{\nu}(y){\cal O}_{\rm ext} 
  \right\rangle
  \label{aawardid}
\end{equation}
with $V^c_{\nu}(y)$ being the isovector vector current.

Eq.~(\ref{aawardid}) is reminiscent of the well-known 
current algebra identities in Minkowski space.
Integrating the axial current
over the boundary $\partial R$ effectively amounts to 
taking the commutator 
of the axial charge with the fields localized in $R$.
In parti\-cu\-lar, the Ward identity eq.~(\ref{aawardid})
simply states that 
the commutator of the axial charge with the axial current 
is equal to the vector current.
Similar Ward identities may be derived for the isovector
vector charge and the whole set of current algebra identities
is thus recovered in euclidean space.

\begin{figure}[t]
\vspace{0.5cm}
\centerline{\epsfig{file=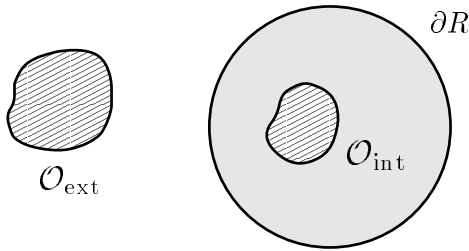,width=6.3cm}}
\vspace{0.3cm}
\caption{Assumed localization regions of the field products 
${\cal O}_{\rm int}$ and ${\cal O}_{\rm ext}$ when deriving the integrated
Ward identity eq.~(\ref{intwardid}).}
\end{figure}

\subsection{Euclidean proof of the Goldstone theorem}

At vanishing quark mass and if chiral symmetry is spontaneously
broken through a non-zero value of $\langle\psibar\psi\rangle$,
the Goldstone theorem asserts that 
there exists an isovector of massless bosons 
(the pions) which 
couple to the pseudo-scalar density $P^a(x)$.
In the following lines we give quick proof of this theorem 
in order to illustrate the power of 
the chiral Ward identities derived above.

Our starting point is the PCAC relation 
\begin{displaymath}
   \langle\partial_{\mu}A^a_{\mu}(x)P^a(0)\rangle=0 
\end{displaymath}
which holds for all $x\neq0$. 
Using Lorentz invariance we then deduce that 
\begin{displaymath}
   \langle A^a_{\mu}(x)P^a(0)\rangle=k{x_{\mu}\over(x^2)^2}
\end{displaymath}
for some constant $k$.
Now if we insert ${\cal O}_{\rm int}=P^a(0)$
and ${\cal O}_{\rm ext}=1$ in 
the integrated Ward identity eq.~(\ref{intwardid}), 
and choose $R$ to be a ball with radius $r$
centred at the origin, the relation
\begin{displaymath}
  \int_{|x|=r}\rmd\sigma_{\mu}(x)\,  
  \langle 
  A^a_{\mu}(x)P^a(0)\rangle
  =-\frac{3}{2}\,\langle\psibar\psi\rangle
\end{displaymath}
is obtained.
The constant $k$ appearing above may thus be calculated and 
one ends up with
\begin{displaymath}
   \langle A^a_{\mu}(x)P^a(0)\rangle=
   -{3\over4\pi^2}\langle\psibar\psi\rangle
   {x_{\mu}\over(x^2)^2}.
\end{displaymath}
We can now see that the current-density correlation function
is long-ranged if
$\langle\psibar\psi\rangle$ does not vanish.
In particular, the energy spectrum does not have a gap
and the correlation function in fact has a particle pole 
at zero momentum.

\subsection{PCAC relation on the lattice}

We now pass to the lattice theory and first discuss the
PCAC relation.
One may be tempted to derive this identity following 
the steps taken in the continuum theory, but 
this would not work out because the chiral variation
of the lattice action cannot be expressed in terms of 
the axial current and density alone
[\Cite{BochicchioEtAl}]. 

We may, however, argue that the position space correlation functions
of properly renormalized lattice fields should converge to the 
corresponding correlation functions in the continuum theory.
In particular, if $(\ar)^a_{\mu}$ and $(\pr)^a$ denote the renormalized
isovector axial current and density, we expect that
\begin{equation}
  \bigl\langle
  \drvtilde{\mu}(\ar)^a_{\mu}(x)\,
  {\cal O}\bigr\rangle
  =2\mr
  \bigl\langle(\pr)^a(x)\,
  {\cal O}\bigr\rangle+\rmO(a)
  \label{PCACb}
\end{equation}
for any product $\cal O$ of fields located at non-zero distances
from $x$ (as before the symbol $\drvtilde{\mu}$ stands for the
average of the forward and backward difference operators).

The size of the error term in eq.~(\ref{PCACb})
directly tells us how strongly chiral symmetry is violated on the lattice.
In particular, in the O($a$) improved theory 
the error should be reduced to order $a^2$
if the coefficients $\csw$ and $\ca$ have their proper values.
This may conversely be taken as a condition to fix these coefficients,
a possibility which will be examined in more detail in
section~5. 

Other applications of the PCAC relation in lattice QCD
include the computation of the 
additive quark mass renormalization $\mc$ [\Cite{paperIII}]
and of the renormalized quark masses
(see ref.~[\Cite{GuptaReview}] for a review and an up-to-date 
list of references).

\exercise{Since the flavour symmetry is unbroken on the lattice,
there exists an exactly conserved isovector vector current.
Deduce the explicit expression for this current}

\subsection{Chiral symmetry \& renormalization}

A well-known theorem states that canonically normalized 
conserved currents need not be 
renormalized, because the possible values of the associated charge
are fixed algebraically. 
On the lattice chiral symmetry is violated and the 
theorem consequently does not apply to the axial current.
In general the renormalization constant $\za$ 
is hence different from~$1$
and needs to be calculated.

\begin{figure}[t]
\vspace{0.5cm}
\centerline{\epsfig{file=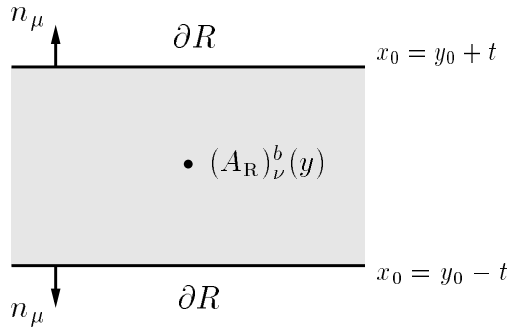,width=6.8cm}}
\vspace{0.3cm}
\caption{Assumed shape of the region $R$ in eq.~(\ref{aav}). 
The vector $n_{\mu}$ denotes the 
outward normal to the boundary $\partial R$.
}
\end{figure}

As has been noted a long time ago,
$\za$ may be determined through the chiral Ward identities
[\Cite{BochicchioEtAl},\Cite{MaianiMartinelli}].
A possible starting point is 
eq.~(\ref{aawardid}) with $R$ being the volume
between two equal-time hyper-planes (see fig.~5).
At zero quark mass and with 
properly normalized currents we then expect that
\begin{eqnarray}
  &&\hspace{-2.0cm}a^3\sum_{x\in\partial R}
  \epsilon^{abc}n_{\mu}
  \langle 
  (\ar)^a_{\mu}(x)
  (\ar)^b_{\nu}(y) {\cal O}_{\rm ext}\rangle
  \nonumber
  \\
  \noalign{\vskip2ex}
  &&\hspace{-1.0cm}\qquad\qquad\qquad =2i
  \left\langle 
  (\vr)^c_{\nu}(y){\cal O}_{\rm ext} 
  \right\rangle
  +\rmO(a)
  \label{aav}
\end{eqnarray}
for any product ${\cal O}_{\rm ext}$ of fields localized outside $R$. 
Note that all fields in these correlation functions
are in space-time regions well separated from each other.
In the O($a$) improved theory, the error term may hence be 
expected to be reduced to order $a^2$.

Since the left-hand side is proportional to the square of $\za$,
and since the axial current does not appear on the other side of 
eq.~(\ref{aav}), it is quite clear that the renormalization constant can be
determined in this way.
We shall not go into any further details here, but
the idea has been
shown to work out in practice and in many cases $\za$ is now known
non-perturbatively [\Cite{ZAa}--\Cite{ZAd}].

Once the renormalization constant of the axial current has been determined,
the chiral Ward identities may be used to relate 
the renormalization constants of fields that belong
to the same chiral multiplet.
The isovector pseudo-scalar density 
$(\pr)^a$ and the isoscalar scalar density $\sr$
transform into each other under chiral rotations and are thus an example
of such a multiplet. 
In this case we would 
replace $A^b_{\nu}(y)$ in eq.~(\ref{aav}) by the scalar density.
The chirally rotated field then appears on the right-hand side 
and by evaluating the correlation functions on both sides of the equation
one is hence able 
to fix the relative normalizations of these fields. 
In principle this procedure generalizes to any chiral multiplet,
but some care should be paid when operator mixing occurs.

\exercise{On the lattice the local isovector vector current 
$V^a_{\mu}=\psibar\dirac{\mu}\frac{1}{2}\tau^a\psi$ is not exactly conserved.
How can one fix the normalization of this current?}

\subsection{Concluding remarks}

Our discussion in this section shows that 
there are many possibilities to make good use of 
chiral symmetry in lattice QCD even though the symmetry
is not preserved on the lattice.
The assumption which has implicitly been made is that 
the lattice corrections to the chiral Ward identities 
are small at the accessible lattice spacings.
In the unimproved theory this is not obviously the case,
because chiral symmetry is violated at order $a$
and these effects can be rather large as we have previously noted.
Further impressive evidence for this will be given below and
the bottom-line then is that 
O($a$) improvement (or maybe some other form of improvement)
should better be applied if one is interested in 
calculating quantities that 
are sensitive to chiral symmetry.

\section{Non-perturbative improvement}

As explained in section~3, 
the coefficients multiplying the various O($a$) counterterms
in the improved theory must be tuned to some 
particular values to achieve the desired improvement.
We now show that 
the coefficient $\csw$ appearing in the improved action, 
eq.~(\ref{improvedaction}), can be determined non-perturbatively
through the requirement that chiral symmetry is 
preserved up to terms of order $a^2$.

The correlation functions that we shall consider
for this purpose are constructed in a finite space-time volume
with Dirichlet boundary conditions in the time direction.
There are good reasons for this choice, but 
these will only become clear later.
Many technical details will be omitted in the following 
to keep the basic argumention transparent.
A complete treatment is given 
in refs.~[\Cite{paperI}--\Cite{paperIII}].

\subsection{Boundary conditions}

The lattices we are going to consider are assumed to have spatial size $L$
and time-like extent $T$, with the time coordinate taking
values in the range $0,a,2a,\ldots,T$. 
We impose periodic boundary conditions in all space directions
and Dirichlet boundary conditions 
at time $x_0=0$ and $x_0=T$.
The dynamical degrees of freedom of the gauge field
are thus the variables $U(x,\mu)$ associated
with the links in the interior of the lattice,
while at the boundaries we require that
\begin{displaymath}
  \left.U(x,k)\right|_{x_0=0}=\rme^{aC},
  \qquad 
  \left.U(x,k)\right|_{x_0=T}=\rme^{aC'},
\end{displaymath}
for all $k=1,2,3$. The matrix 
$C$ is taken to be constant diagonal, 
\begin{displaymath}
  C={i\over L}\pmatrix{\phi_1 & 0      & 0      \cr
                       0      & \phi_2 & 0      \cr
                       0      & 0      & \phi_3 \cr},
  \qquad
  \phi_1+\phi_2+\phi_3=0,
\end{displaymath}
and $C'$ is similarly given in terms of another set of angles $\phi'_{\alpha}$. 

In the case of the quark field only half of the Dirac 
components are fixed at the boundaries.
The quark propagator is then the solution of a well-posed boundary
value problem. 
Explicitly, if we introduce
the projectors $P_{\pm}=\frac{1}{2}(1\pm\dirac{0})$,
the boundary conditions are
\begin{displaymath}
  P_{+}\psi(x)|_{x_0=0}=\rho({\bf x}),
  \qquad
  P_{-}\psi(x)|_{x_0=T}=\rhoprime({\bf x}),
\end{displaymath}
where $\rho$ and $\rhoprime$ are some externally given 
(anti-commuting) fields.
The corresponding boundary conditions on 
the anti-quark field are
\begin{displaymath}
  \psibar(x)P_{-}|_{x_0=0}=\rhobar({\bf x}),
  \qquad
  \psibar(x)P_{+}|_{x_0=T}=\rhobarprime({\bf x}).
\end{displaymath}
The field components $\psi(x)$ and $\psibar(x)$ 
at times $0<x_0<T$ remain unconstrained and represent
the dynamical part of the quark and anti-quark fields.

\exercise{Consider the free quark theory in the continuum limit
with boundary conditions as specified above. Solve the 
Dirac equation for arbitrary smooth boundary values $\rho,\rho'$ and prove
that the solution is unique}

\subsection{Correlation functions in finite volume}

The lattice action in finite volume is essentially given by the 
expressions that we have written down 
in the preceeding sections for the infinite volume theory.
In the case of the O($a$) counterterm in eq.~(\ref{improvedaction}),
for example, the only modification is that the sum
is restricted to all points $x$ in the interior of the lattice.

Expectation values of products $\cal O$ of local fields
are now obtained in the usual way through the functional integral.
The integration is performed at fixed boundary values, i.e.~the 
integration variables are the unconstrained 
dynamical field variables residing
in the interior of the lattice.
In particular, the only dependence on the boundary values
$C,C',\rho,\rhobar,\rhoprime$ and $\rhobarprime$
arises from the action.

An interesting option at this point is to 
include derivatives with respect to the quark and anti-quark boundary values
in the field product $\cal O$. The operator
\begin{equation}
  {\cal O}^a=
  -a^6\sum_{\bf y,z}
  {\delta\over\delta\rho({\bf y})}\dirac{5}\frac{1}{2}\tau^a
  {\delta\over\delta\rhobar({\bf z})}
  \label{source}
\end{equation}
is an example of such a product. In the functional integral
the derivatives act on the weight factor $\rme^{-S}$ and have the
effect of inserting certain combinations of the dynamical fields
localized near the boundaries of the lattice.
More precisely, one finds that in nearly all respects the derivatives
$-\delta/\delta\rho({\bf x})$ 
and $\delta/\delta\rhobar({\bf x})$ 
behave like an anti-quark and a quark field at time $x_0=0$.

\begin{figure}[t]
\vspace{0.5cm}
\centerline{\epsfig{file=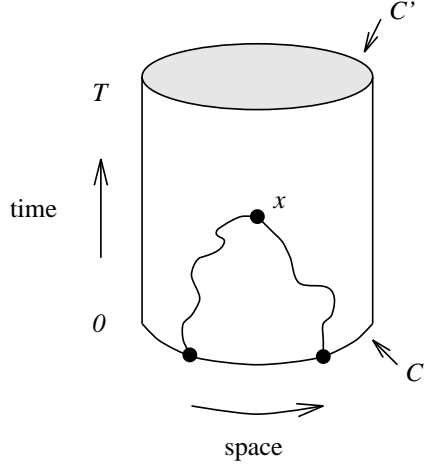,width=5.5cm}}
\vspace{0.3cm}
\caption{
Sketch of the space-time manifold on which the lattice theory
is set up. $C$ and $C'$ are the boundary values of the gauge field.
The irregular lines represent
the trajectory of a quark anti-quark pair, which is
created at time $0$ through the operator ${\cal O}^a$
[eq.~(\ref{source})].
}
\end{figure}
 
For illustration let us consider 
the correlation function
\begin{equation}
  \langle A_{\mu}^a(x)\op{}^a\rangle=
  \left\{{1\over{\cal Z}}\int_{\rm fields}A^a_{\mu}(x)
  \op{}^a\,\rme^{-S}\right\}_{\rho=\rhobar=\rhoprime=\rhobarprime=0}.
  \label{corrfunc}
\end{equation}
Since one sums
over the positions $\bf y,z$ in eq.~(\ref{source}), 
the operator creates a quark and an anti-quark at time zero with 
vanishing spatial momentum.
The correlation function 
then is proportional to the probability amplitude that 
the quark anti-quark pair propagates to the interior of the 
space-time volume and that it annihilates at the point $x$
(see fig.~6). As usual such pictures have 
an exact meaning if the quark lines are thought to represent  
quark propagators at the current gauge field [\Cite{paperII}].

\subsection{How large are the chiral symmetry violations?}

In principle the error term on the right-hand side of the 
PCAC relation 
eq.~(\ref{PCACb}) provides an estimate of
the size of chiral symmetry violation in lattice QCD.
The renormalization factors in the expressions for
the renormalized improved axial current
and the associated pseudo-scalar density,
\begin{eqnarray*}
  (\ar)^a_{\mu}&=&\za(1+\ba a\mq)
  \bigl\{A^a_{\mu}+\ca a\drvtilde{\mu}P^a\bigr\}, \\
  \noalign{\vskip1.5ex}
  (\pr)^a&=&\zp(1+\bp a\mq)P^a,
\end{eqnarray*}
are however not known at this point and
a straightforward calculation of the error term
is hence not possible.

Now let us define 
an unrenormalized current quark mass through
\begin{equation}
  m={\bigl\langle
  \bigl\{\drvtilde{\mu}A^a_{\mu}+
  \ca a\drvstar{\mu}\drv{\mu}P^a\bigr\}
  {\cal O}^a\bigr\rangle
  \over
  2\bigl\langle P^a{\cal O}^a\bigr\rangle},
  \label{mcurrent}
\end{equation}
where ${\cal O}^a$ is the operator introduced above.
The PCAC relation then implies
\begin{equation}
  m={\zp(1+\bp a\mq)\over\za(1+\ba a\mq)}\,\mr+\rmO(a).
  \label{mmr}
\end{equation}
The renormalization constants and the renormalized mass do not depend
on the kinematical parameters such as the time $x_0$ at which the 
axial current is inserted or the boundary values $C$ and $C'$
of the gauge field. Changing these parameters basically means
to probe the PCAC relation in different ways.
So if we consider two configurations of the kinematical
parameters at the same point $(g_0,am_0)$ in the bare parameter space,
and if $m_1$ and $m_2$ denote the 
associated values of $m$, it follows that 
\begin{equation}
  m_1-m_2=\rmO(a).
  \label{massdiff}
\end{equation}
By calculating $m_1$ and $m_2$ we thus obtain a direct check on 
the size of the lattice effects in the PCAC relation.

In the improved theory one expects that the differences in the calculated
values of $m$ are reduced to order $a^2$. 
This may not be totally obvious,
because improvement has only been discussed in the context of 
the infinite volume theory.
The boundaries at $x_0=0$ and $x_0=T$ indeed require
the introduction of further O($a$) counterterms,
but since the PCAC relation is locally derived,
it can be shown that these affect the relation
at order $a^2$ only [\Cite{paperI}].

\subsection{Tests of chiral symmetry at tree-level of perturbation theory}

At vanishing coupling and if $C=C'=0$, the quarks propagate
freely and the current quark mass $m$ is hence expected to coincide
with the bare mass $m_0$ up to lattice effects.
A straightforward calculation confirms this and one also finds that 
$m$ is independent of the time $x_0$ at which the 
axial current is inserted as it should be if lattice effects 
are negligible.

\begin{figure}[t]
\vspace{0.0cm}
\centerline{\epsfig{file=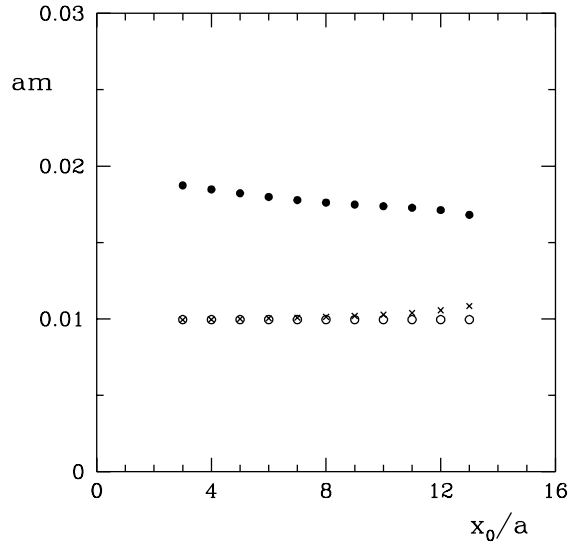,width=10.0cm}}
\vspace{-1.2cm}
\caption{
Plot of $am$ on a $16\times8^3$ lattice with $am_0=0.01$
at tree-level of perturbation theory.
The boundary values of the gauge field are zero (open circles)
or as given by eq.~(\ref{bfielda}). Full circles and crosses 
correspond to $\csw=0$ and $\csw=1$ respectively.
}
\end{figure}
 
For non-zero boundary values $C$ and $C'$
the situation changes and large lattice effects are 
observed in the unimproved theory.
To illustrate this we consider a $16\times8^3$ lattice and set
\begin{equation}
  (\phi_1,\phi_2,\phi_3)=
  \frac{1}{6}\left(-\pi,0,\pi\right),
  \qquad
  (\phi'_1,\phi'_2,\phi'_3)=
  \frac{1}{6}\left(-5\pi,2\pi,3\pi\right).
  \label{bfielda}
\end{equation}
As shown in fig.~7 the corresponding values of $m$
strongly deviate from the free quark value
and they are also far from being independent of $x_0$.
These effects almost completely disappear when the improved
action is used with $\csw=1$ (which is the proper value to this order of 
perturbation theory).
Improvement thus works very well and it is possible to prove that 
the residual lattice effects seen in fig.~7 are of order $a^2$ as expected.

The outcome of these calculations can be readily understood
if we note that the boundary values $C$ and $C'$ induce a 
background gauge field in the space-time volume, equal to 
the least action configuration interpolating between $C$ and $C'$.
For the boundary values chosen above,
the background field can be shown to be 
a constant colour-electric field given by
\begin{equation}
  U(x,\mu)=\rme^{aB_{\mu}(x)},
  \qquad
  B_0(x)=0,
  \qquad
  B_k(x)=C+\left(C'-C\right)x_0/T.
  \label{bfieldb}
\end{equation}
Now since the lattice effects of order $a$ can be described by an
effective Pauli term, their magnitude is proportional to the 
strength of the background field and the tests
that we have performed are hence directly probing for these effects.

\exercise{Check that the PCAC relation is violated by O($a^2$) effects
only in the free quark theory on an infinite lattice}

\begin{figure}[t]
\vspace{0.0cm}
\centerline{\epsfig{file=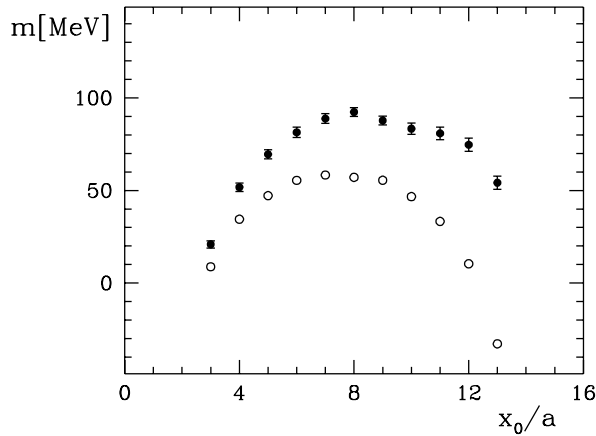,width=10.0cm}}
\vspace{-2.0cm}
\caption{Unrenormalized current quark mass $m$ 
on a $16\times8^3$ lattice with spacing $a=0.05$ fm,
calculated using the unimproved action.
Open and full circles correspond to zero and non-zero boundary values
of the gauge field. 
}
\end{figure}

\subsection{Computation of $\csw$}

Without improvement the lattice corrections to the PCAC relation
are rather large also at the non-perturbative level.
This is made evident through fig.~8, where
the results of a numerical simulation of the unimproved theory
in the quenched approximation are plotted.
In this calculation the bare mass $m_0$ has been set to some 
value close to $\mc$ and the bare coupling has been chosen so that 
$6/g_0^2=6.4$, which corresponds to a lattice spacing of 
about $0.05$ fm.
Again one observes a strong dependence of the calculated
values of $m$ on the background field and also on $x_0$.

\begin{figure}[t]
\vspace{0.0cm}
\centerline{\epsfig{file=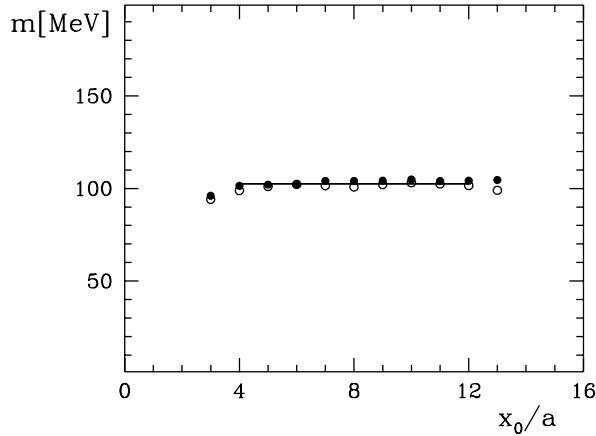,width=10.0cm}}
\vspace{-2.0cm}
\caption{Same as fig.~8, but using the improved action and the improved
axial current with $\csw=1.60$ and $\ca=-0.027$.
The horizontal line is drawn to guide the eye.
}
\end{figure}

We may now attempt to adjust the coefficients $\csw$ and $\ca$ 
so as to cancel these effects.
This works remarkably well and it is possible to 
bring almost all points to a well-defined plateau (see fig.~9).
Similar results are obtained for larger lattices, other boundary
values and so on. In general improvement is very efficient,
at least for $a\leq0.1$ fm, and with properly chosen 
$\csw$ and $\ca$ the remaining violations of chiral symmetry 
are small.

\subsection{Results and further developments}

Proceeding in this way the coefficient $\csw$
has been determined in quenched QCD 
[\Cite{paperIII},\Cite{SCRIa}] 
and now also in full QCD with a doublet of dynamical quarks
[\Cite{KarlRainer}].
In the case of full QCD the range of couplings covered 
extends from small couplings, where contact is made with the one-loop formula
eq.~(\ref{cswoneloop}), up to couplings corresponding to 
$6/g_0^2\simeq5.2$, where the lattice spacing is around $0.1$ fm
according to some preliminary studies (see fig.~10). 
The non-perturbative values of $\csw$ deviate considerably
from the one-loop result at the larger couplings,
which is the relevant range for 
calculations of the hadron masses and other other properties
of these particles.

\begin{figure}[t]
\vspace{0.0cm}
\centerline{\epsfig{file=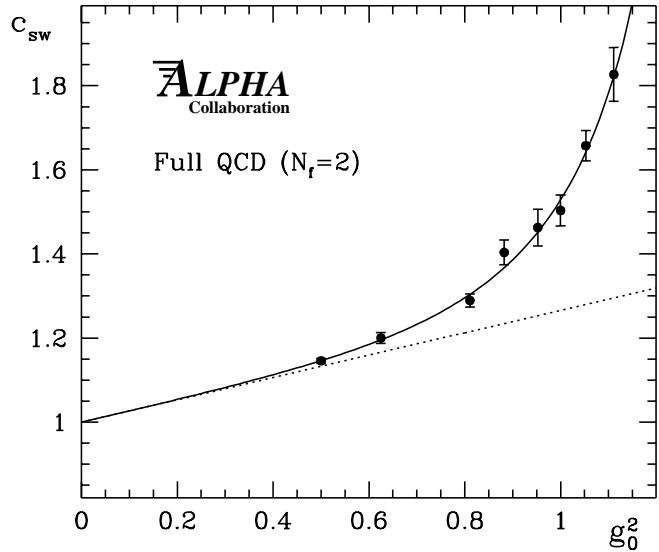,width=12.0cm}}
\vspace{-2.5cm}
\caption{Simulation results for $\csw$ in QCD with a doublet
of dynamical quarks [\Cite{KarlRainer}]. 
The solid curve represents the fit eq.~(\ref{cswfit})
and the dotted line the one-loop formula eq.~(\ref{cswoneloop}).
}
\end{figure}

The data shown in fig.~10 are well represented by the rational expression
\begin{equation}
  \csw={1-0.454\,g_0^2-0.175\,g_0^4+0.012\,g_0^6+0.045\,g_0^8
  \over
  1-0.720\,g_0^2}.
  \label{cswfit}
\end{equation}
At the level of the numerical accuracy that one has been able
to reach, this fit is no worse than the simulation data
and it may at this point just as well 
be taken as the definition of $\csw$.
An important remark in this connection is that
different ways to calculate $\csw$, using various background fields etc.,
yield results that differ by terms of order $a$.
There is nothing fundamentally wrong with this. 
The systematic uncertainty in $\csw$ 
merely reflects the fact that O($a$) improvement is an 
asymptotic concept valid up to higher-order terms.

Once $\csw$ is known non-perturbatively, one may attempt to 
calculate the coefficients of other O($a$) correction terms
using similar techniques. 
So far these calculations have been limited
to quenched QCD and results have been published
for $\ca$ [\Cite{paperIII}], the 
O($a$) counterterm needed to improve the isovector vector current
[\Cite{MarcoRainer}] and various $b$-coefficients 
[\Cite{ZAd},\Cite{bmNonpert}]. Some of these
appear to be less accessible 
and new ideas will be required 
to be able to compute them [\Cite{MartinelliEtAlIII}].

\section{Non-perturbative renormalization}

One of the fundamental problems in QCD is to 
establish the connection between the low-energy sector and the
perturbative regime of the theory.
There are many instances where this is required, 
a particularly obvious case being
hadronic matrix elements of operators, whose
normalization is specified at high energies through the 
$\msbar$ scheme of dimensional regularization.
In the present section we address the issue
in the context of lattice QCD and
briefly describe some of the
techniques that have been proposed to resolve it.

\subsection{Example}

Let us first discuss one of the standard ways
to compute the running quark masses in lattice QCD.
The need for non-perturbative renormalization will
then become clear. Any details not connected with this
particular aspect of the calculation are omitted.

A possible starting point to obtain the sum 
$\mbar_{\rm u}+\mbar_{\rm s}$ of the up and 
the strange quark masses is the PCAC relation
\begin{displaymath}
  \mK^2f_{\lower1.0pt\hbox{\sixrm K}}=(\mbar_{\rm u}+\mbar_{\rm s})
  \langle0|\,\bar{u}\dirac{5}s\,|K^{+}\rangle
\end{displaymath}
(lattice effects are ignored here).
Since the kaon mass $\mK$ and the decay constant $\fK$ are 
known from experiment, it suffices to evaluate the matrix element
on the right-hand side of this equation.
On the lattice one first computes the matrix element of 
the bare operator $(\bar{u}\dirac{5}s)_{\lat}$
and then multiplies the result with the renormalization factor $\zp$
relating $(\bar{u}\dirac{5}s)_{\lat}$ to the renormalized
density $\bar{u}\dirac{5}s$.

\renewcommand{\arraystretch}{1.3}
\begin{table}[t]
\caption{Recent results for $\mbar_{\rm s}$
(quenched QCD, $\msbar$ scheme at $\mu=2$ GeV)}
\vspace{0.4cm}
\begin{center}
\footnotesize
\begin{tabular}{|l|l|}
\hline
\multicolumn{1}{|c|}{$\mbar_{\rm s}$ [MeV]}   
& \multicolumn{1}{c|}{reference} \\
\hline
\hspace{0.5cm}$122(20)$\hspace{0.5cm}  
& Allton et al.~(APE collab.)~\cite{QM_AlltonEtAl} \\
\hspace{0.5cm}$112(5)$                      
& G\"ockeler et al.~(QCDSF collab.)~\cite{QCDSFa} \\
\hspace{0.5cm}$111(4)$                      
& Aoki et al.~(CP-PACS collab.)~\cite{QM_CPPACS} \\
\hspace{0.5cm}$\phantom{0}95(16)$           
& Gough et al.~\cite{QM_GoughEtAl}  \\
\hspace{0.5cm}$\phantom{0}88(10)$           
& Gupta \& Bhattacharya~\cite{QM_Gupta} \\
\hline
\end{tabular}
\end{center}
\end{table}
\renewcommand{\arraystretch}{1.0}

Some recent results for the strange quark mass 
obtained in this way or in similar ways are listed in Table~1
(further results can be found
in the review of Bhattacharya and Gupta~\cite{GuptaReview}).
The sizeable differences among these numbers
have many causes.
An important uncertainty arises from the fact that, 
in one form or another, the one-loop formula
\begin{displaymath}
  \zp=1+{g_0^2\over4\pi}\left\{(2/\pi)\ln(a\mu)+k\right\}
      +\rmO(g_0^4)
\end{displaymath}
has been used to compute the renormalization factor,
where $\mu$ denotes the
normalization mass in the $\msbar$ scheme and $k$ a calculable constant
that depends on the details of the lattice regularization.
Bare perturbation theory has long been known to be unreliable 
at the couplings of interest and without further insight
it is clearly impossible to reliably assess the error on 
the so calculated values of $\zp$.
Evidently this difficulty would go away if we were able
to compute the renormalization constant non-perturbatively.

\subsection{Hadronic renormalization schemes}

We now proceed to introduce the notion of a
hadronic renormalization scheme and 
shall then be able to formulate the problem
of non-perturbative renormalization in more physical terms.

In a hadronic scheme the renormalization conditions are 
imposed at low energies by requiring a set of physical quantities,
such as the hadron masses and suitable hadronic matrix elements,
to assume prescribed values.
We may, for example, take the pion decay constant $\fpi=132$~MeV
as the basic reference scale and fix the quark masses 
through the dimensionless ratios
\begin{displaymath}
  \mpi/\fpi,\,\mKplus/\fpi,\,\mKnull/\fpi,\,\ldots
\end{displaymath}
On the lattice 
the combination $a\fpi$ then becomes to a function of $g_0$ alone
and a variation in the coupling 
thus amounts to changing the lattice spacing
at constant $\fpi$ (and fixed quark masses).

The normalization of local fields can be fixed similarly
by requiring the renormalized fields $\Or$
to satisfy
\begin{displaymath}
  \langle\,f\,|\Or|\,i\,\rangle=
  \hbox{prescribed value}
\end{displaymath}
for some hadronic states $|\,i\,\rangle$ and $|\,f\,\rangle$.
If there is operator mixing
further normalization conditions must be imposed 
to fix the finite parts of the mixing coefficients.

After these preparations we can now 
clearly say what non-perturbative renormalization means.
Basically what we require is to match a given hadronic scheme with
the $\msbar$ scheme of
dimensional regularization (or any other perturbative scheme).
This amounts to calculating the running coupling and quark masses
at high energies
as well as the renormalization constants 
$X_{\cal O}$ that one needs to convert from the renormalized 
operators $\Or$ to the corresponding fields
the $\msbar$ scheme.

In principle such a computation is straightforward. 
The key observation is that 
all physical amplitudes are uniquely determined 
functions of the external momenta and the lattice spacing
once a definite hadronic scheme has been adopted.
We can then consider a set of correlation functions of local
fields at large momenta and arrange the momentum cutoff
$\pi/a$ to be far above these scales.
Lattice effects are negligible under these conditions
and by comparing the exact correlation functions with
their perturbation expansion in the $\msbar$ scheme one 
is hence able to extract the running parameters and 
renormalization constants at normalization scales
$\mu$ given in units of $\fpi$ for example.

A subtle technical point here is that the $\msbar$ scheme is 
only defined to every finite order of perturbation theory.
The statement that the running coupling etc.~can
be calculated has, therefore, a precise meaning 
only in the limit $\mu/\fpi\to\infty$ where higher-order 
corrections become negligible.

\exercise{How precisely are different hadronic schemes related 
to each other? Convince yourself that it suffices to solve the 
non-perturbative renormalization problem for one particular
hadronic scheme}

\subsection{Meanfield improved perturbation theory}

One of the attempts to solve the non-perturbative
renormalization problem goes back to 
an observation of Parisi [\Cite{Parisi}], who noted that 
the often poor convergence of the bare perturbation expansion
may be due to the appearance of so-called tadpole diagrams.
Meanfield theory then led him to suggest that these diagrams
can be resummed by replacing the coupling $\alpha_0=g_0^2/4\pi$ 
through $\alphaP=\alpha_0/P$, 
where $P=1-\frac{1}{3}g_0^2+\ldots$ denotes the plaquette expectation value.
Manipulations of this sort have
later been discussed in greater detail 
by Lepage and Mackenzie [\Cite{Lepenzie}]
and their recipes are now widely used.

For illustration let us consider the series 
\begin{equation}
  \alphaMSbar(\mu)=
  \alpha_0+d_1(a\mu)\alpha_0^2+d_2(a\mu)\alpha_0^3+\ldots,
  \label{AlphaExpansionA}
\end{equation}
which one obtains by calculating some correlation functions at 
large momenta using bare perturbation theory and comparing with 
the expansion in the $\msbar$ scheme. 
For the standard Wilson action the expansion coefficients
are known up to two loops
[\Cite{Hasenfratzsq}--\Cite{TwoLoopII}], the result in
quenched QCD being
\begin{eqnarray*}
  d_1(a\mu)&=&5.8836-{11\over2\pi}\ln(a\mu),
  \\
  \noalign{\vskip0.5ex}  
  d_2(a\mu)&=&[d_1(a\mu)]^2+8.7907-{51\over4\pi^2}\ln(a\mu).
\end{eqnarray*}
Note that we do not require $\mu$ to be much smaller than 
the momentum cutoff $\pi/a$. When deriving the expansion
eq.~(\ref{AlphaExpansionA}),
one only needs to know that the external momenta 
entering the lattice correlation functions that one has chosen to work out
are well below this scale. The normalization mass, on the other hand,
only appears in the continuum theory and can be arbitrarily large.

To apply these formulae
one selects some value of the bare coupling 
and determines the lattice spacing in the chosen hadronic scheme.
This step involves the calculation of meson masses etc.~and is
hence performed using numerical simulations.
Knowing the lattice spacing in physical units and the bare coupling,
the right-hand side of eq.~(\ref{AlphaExpansionA}) can be evaluated
straightforwardly for any given $\mu$.
On the presently accessible lattices, 
$g_0$ is around $1$ and $1/a$ is a few GeV at most.
So if we set $\mu$ to some reasonable value, say $10$ GeV, 
the one- and two-loop terms in the expansion make a contribution 
roughly equal to $35\%$ and $17\%$ of the leading term.
These corrections are not small and there is, therefore, little
reason to trust the results that one obtains.

Now if we take Parisi's coupling $\alphaP$ as the expansion parameter,
the terms in the resulting series tend to be significantly reduced.
We may, for example, adjust 
the scale factor $a\mu$ so that the one-loop term 
vanishes and the two-loop correction in the resulting expansion
\begin{displaymath}
  \alphaMSbar(\mu)=\alphaP+2.185\times\alphaP^3+\ldots,
  \qquad \mu=2.633\times 1/a,
\end{displaymath}
then is indeed rather small
($\alphaP$ is around $0.13$ at the bare couplings of interest).
As a consequence one may be more confident in applying
this form of the perturbation expansion,
although it remains unclear how the truncation error can be 
estimated in a reliable manner.

\subsection{Intermediate renormalization}

An interesting method to compute renormalization factors
that does not rely on bare perturbation theory has been
proposed by Martinelli et al.~[\Cite{IR_Martinelli}].
The idea is to proceed in two steps, first matching
the lattice with an intermediate momentum subtraction (MOM) scheme 
and then passing to the $\msbar$ scheme.
The details of the intermediate MOM scheme 
do not influence the final results and are of only practical
importance. One usually chooses the Landau gauge and imposes
normalization conditions on the propagators and the 
vertex functions at some momentum~$p$.
In the case of the pseudo-scalar density, for example,
the renormalization constant $\zpmom$ is defined through

\vbox{
\vspace{1.0cm}
\begin{equation}
=\;\ztwomom/\zpmom\;\times
\end{equation}}
\vbox{
\vspace{-1.35cm}
\hbox{\epsfxsize=6.0cm\hspace{3.8cm}\epsfbox{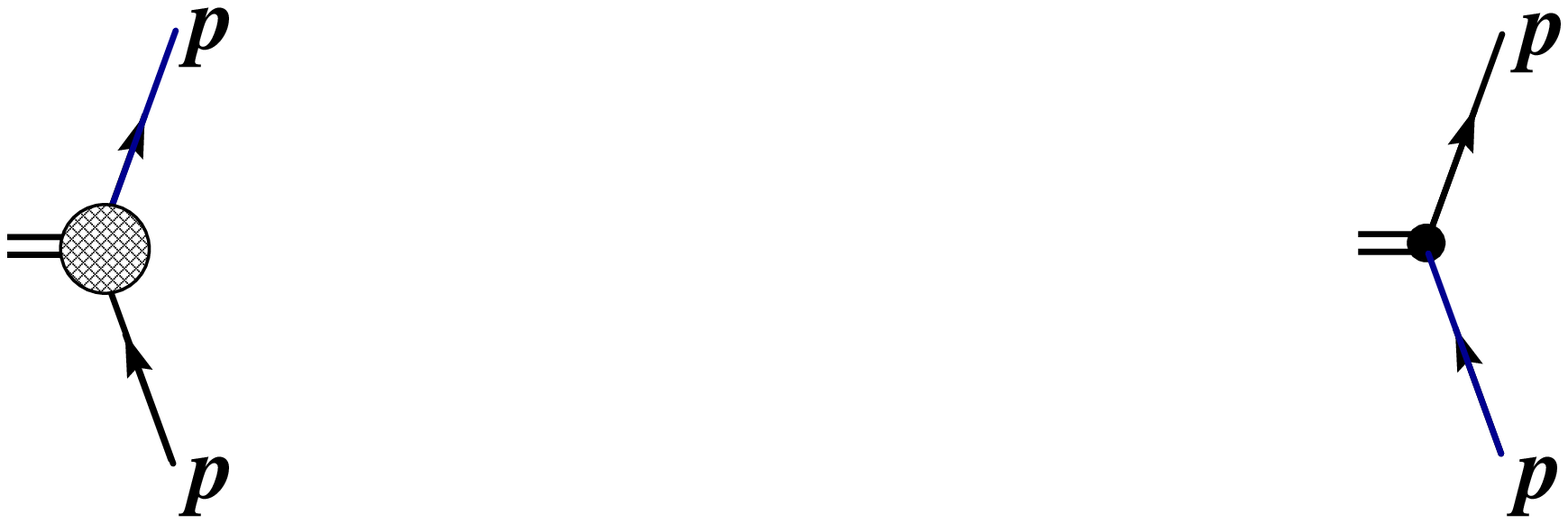}}
\vspace{0.3cm}}

\noindent
where $\ztwomom$ denotes the quark wave function renormalization constant
and the diagrams represent the full and the bare vertex function 
associated with this operator. 

On a given lattice and for a range of momenta,
the quark propagator and the full vertex function
can be computed using numerical 
simulations~[\Cite{IR_Martinelli},\Cite{IR_Giusti}].
$\ztwomom$ and $\zpmom$ are thus obtained non-perturbatively.
The total renormalization factor relating the lattice normali\-zations with 
the $\msbar$ scheme is then given by
\begin{equation}
   \zp(g_0,a\mu)=\zpmom(g_0,ap)\xpmom(\gmsbar,p/\mu),
\end{equation}
with $\xpmom$ being the finite renormalization constant required to match
the MOM with the $\msbar$ scheme. $\xpmom$ is known
to one-loop order of renormalized perturbation theory 
and could easily be worked out to two loops.

While this method avoids the use of bare perturbation theory,
it has its own problems, the most important being 
that the momentum $p$ should be significantly
smaller than $\pi/a$ to suppress the lattice effects,
but not too small as otherwise one may not be confident to 
apply renormalized perturbation theory to compute $\xpmom$.
It is not totally obvious that these conditions can be met on the 
currently accessible lattices~[\Cite{IR_Crisafulli},\Cite{IR_Goeckeler}]
and careful studies should be made to check this.

\subsection{Recursive finite-size technique}

The computational strategies for non-perturbative renormalization
discussed so far assume that all relevant physical scales
can be accommodated on a single lattice which is sufficiently small
for the required calculations to be performed using numerical simulations.
As a consequence the energy range where the hadronic
scheme can be matched with perturbation theory tends to be 
rather narrow and it is then not easy to control the 
systematic errors.

As has been noted some time ago~[\Cite{FSTa}], 
this difficulty can be overcome
by simulating a sequence of lattices
with decreasing lattice spacings.
Any single lattice only covers a limited range of distances,
but through the use of a finite-volume renormalization scheme
it is possible to match subsequent lattices. 
Effectively one thus constructs a non-perturbative renormalization group.
In a few steps it is then possible to reach much higher energies
than would otherwise be possible.

The technique is generally applicable and does not require
any uncontrolled approximations 
to be made~[\Cite{FSTb}--\Cite{FSTe}].
Perhaps the greatest difficulty is that 
the basic strategy does not easily disclose itself.
One also needs to perform extensive numerical simulations 
to set up the non-perturbative renormalization group
and there is usually a significant amount of 
analytical work to be done.

In the following section
we discuss the physical picture underlying 
the technique and introduce some of the key notions.
The method itself will then be explained in section~8,
taking the computation of the running coupling in quenched
QCD as an example.

\section{QCD in finite volume and the femto-universe}

In quantum field theory the physical information is 
encoded in the correlation functions of local operators
and these are hence the primary quantities to consider.
From statistical mechanics one knows, however, that 
certain properties of the system can often be determined more easily
by studying its behaviour in
finite volume. The calculation of critical exponents is 
a classical case where such finite-size techniques are
being applied.

The questions one would like to answer in QCD are 
not the same as in statistical mechanics,
but the general idea to probe the system through 
a finite volume proves to be fruitful here too.
In this section our aim mainly 
is to provide a qualitative understanding
of what happens when the volume is decreased.
Unless stated otherwise, periodic boundary conditions
are assumed and the lattice spacing is taken to be 
much smaller than the relevant physical scales
so that lattice effects can be ignored.

\subsection{Physical situation from large to small volumes}

Let us first consider the case where the spatial extent $L$
of the lattice 
is significantly greater than the typical size of the hadrons
(box~(a) in fig.~11). Single hadrons are practically unaffected
by the finite volume under these conditions except that their momenta
must be integer multiples of $2\pi/L$.
For multi-particle states the situation is a bit more complicated,
because the particles cannot get very far away from each other.
Two-particle energy eigenstates, for example,
really describe stationary scattering processes.
If there are no resonances the corresponding energy values 
differ from the spectrum 
calculated for non-interacting particles
by small amounts proportional to $1/L^3$
[\Cite{Vola}--\Cite{Vold}].

\begin{figure}[t]
\vspace{0.0cm}
\centerline{\epsfig{file=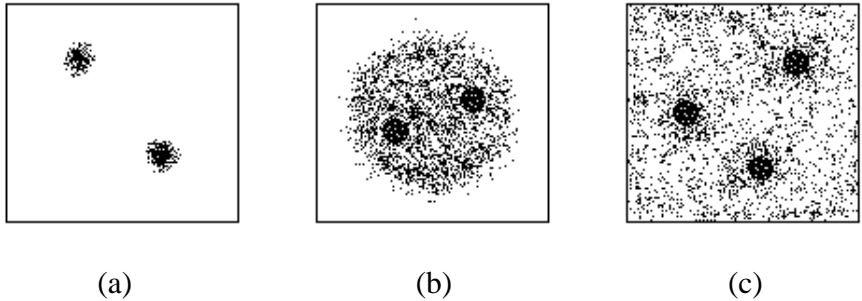,width=12.0cm}}
\vspace{0.0cm}
\caption{Pictures illustrating various physical situations in 
finite volume.
(a) Hadrons in a large volume, (b) a ${\rm q}\bar{\rm q}$ 
meson in a box of size $L\simeq2$ fm, 
and (c) quarks in the femto-universe.
}
\end{figure}

When $L$ is $2$ fm or so, a single hadron barely fits 
into the box and the virtual pion cloud around 
the particle may be slightly distorted
(box~(b) in fig.~11). 
In particular, via the periodic boundary conditions it
is possible that a pion is exchanged ``around the world".
One of the consequences of this effect is that the 
masses of the hadrons are shifted relative to their
infinite volume values by terms of 
order $\rme^{-\mpi L}$~[\Cite{Vola}]. 

At still smaller values of $L$, the quark wave functions of 
the enclosed hadrons are squeezed and 
one observes rapidly increasing volume effects.
Eventually when $L$ drops below the confinement radius,
say $L\leq0.5$ fm,
the quarks are liberated and hadrons cease to exist
(box~(c) in fig.~11).
QCD in such small volumes is referred to as the femto-universe [\Cite{AFa}].
Evidently physics is very different in this little world from what 
one is used to. The chosen boundary conditions
play an important r\^ole, for example, and the energy levels 
are separated by large gaps that increase proportionally to $1/L$.

\exercise{Work out the volume dependence of the particle mass
in the massive $\lambda\phi^4$ theory to first order in $\lambda$.
Check that the volume effects vanish exponentially at large~$L$}

\subsection{Asymptotic freedom and the limit $L\to0$}

Now if we go to the extreme and allow the box size to become  
arbitrarily small, we are entering the short distance regime of QCD
and thus expect that the effective gauge coupling
scales to zero according to the perturbative renormalization group.

To make this a bit more explicit, let us consider the pure gauge theory
on a lattice of infinite extent in the time direction
and twisted periodic or Dirichlet boundary conditions in the space 
directions. The reason for this exotic choice is that perturbation theory
is much more complicated if periodic boundary conditions are imposed
due to the appearance of physical zero modes in the gluon action
[\Cite{AFb}--\Cite{AFi}].
In this theory 
an effective gauge coupling may be defined through
\begin{displaymath}
  \alphaqqbar=\Bigl\{\frac{3}{4}r^2\Fqqbar(r,L)\Bigr\}_{r=L/2},
\end{displaymath}
where $\Fqqbar(r,L)$ denotes the force between static 
quarks at distance $r$ in finite volume.
Note that $L$ is the only external physical scale 
on which this coupling depends.

Perturbation theory at finite $L$ is more complicated
than in infinite volume, but with the chosen boundary conditions
no fundamental difficulties are encountered 
when deriving the Feynman rules.
It is possible to set up perturbation theory 
directly in the continuum theory using dimensional regularization.
Alternatively one may choose to work on the lattice and to convert
to the $\msbar$ scheme later (if so desired).
Proceeding either ways the expansion 
\begin{equation}
  \alphaqqbar
  =\alphaMSbar(\mu)+
  \left[\left(11/2\pi\right)\ln(\mu L)+k\right]\alphaMSbar(\mu)^2+\ldots
  \label{AlphaExp}
\end{equation}
may be derived and if we set $\mu=1/L$ in this equation 
it is immediately clear that $\alphaqqbar$ behaves like a 
running coupling at scale $L$.

In full QCD with any number quarks,
essentially the same arguments apply
and the important conclusion then is 
that QCD is perturbatively soluble at sufficiently small $L$.
Eq.~(\ref{AlphaExp}) also provides a link 
between the femto-universe and the theory in infinite volume.
In particular, if we manage to compute
$\alphaqqbar$ for some box size given in physical units, 
we could use the relation to determine the $\msbar$ coupling at 
this scale. 
This may be surprising at first sight,
but the equation only reflects the fact that the basic lagrangian
is the same independently of the physical context.

\subsection{Finite-volume renormalization schemes}

It should be quite clear at this point that the box size $L$
may be regarded as a reference scale similar to the 
normalization mass in the $\msbar$ scheme.
Renormalization schemes that scale with $L$ are in fact 
easily constructed by choosing some particular boundary conditions
and by scaling all dimensionful external parameters
entering the renormalization conditions
with the appropriate power of $L$.
In particular, if the time-like extent $T$ of the lattice is finite,
the ratio $T/L$ should be set to some definite value.

The coupling $\alphaqqbar$ introduced above 
complies with this general definition and 
there are many ways to set up
finite-volume renormalization conditions for 
the local operators of interest.
The normalization of the isovector pseudo-scalar density, for example, 
can be fixed through
\begin{displaymath}
  a^6\sum_{\bf x,y}
  \left.\langle(\pr)^a(x)(\pr)^a(y)\rangle\right|_{x_0-y_0=L/2}=
  \hbox{constant},
\end{displaymath}
where the constant should be adjusted so that
$\pr$ has the standard normalization at tree-level of perturbation 
theory. If zero modes are excluded by the boundary conditions,
one can then show that  
\begin{displaymath}
  \pr=
  \left\{1+\left[-(2/\pi)\ln(\mu L)+k\right]\alphaMSbar(\mu)+\ldots
  \right\}\pMSbar(\mu)
\end{displaymath}
(with another constant $k$).
From the point of view of perturbation theory,
finite-volume renormalization schemes are hence as good
as any other scheme.

\exercise{Invent a finite-volume renormalization condition for the 
quark masses}

\section{Computation of the running coupling}

Non-perturbative renormalization is required
in many places in lattice QCD, but the calculation of 
the running coupling is certainly one of the most 
important cases to consider.
As already mentioned in section~6,
finite-size techniques allow one to 
trace the evolution of the coupling from low to high energies
and we would now like to describe this in some more detail.
Although the method is generally applicable, 
attention will here be restricted to the pure SU(3) gauge theory.
This automatically includes quenched QCD, since the
renormalization of the coupling is independent of the valence quarks,
but in full QCD a separate calculation will be required.

\subsection{Strategy}

\begin{figure}[t]
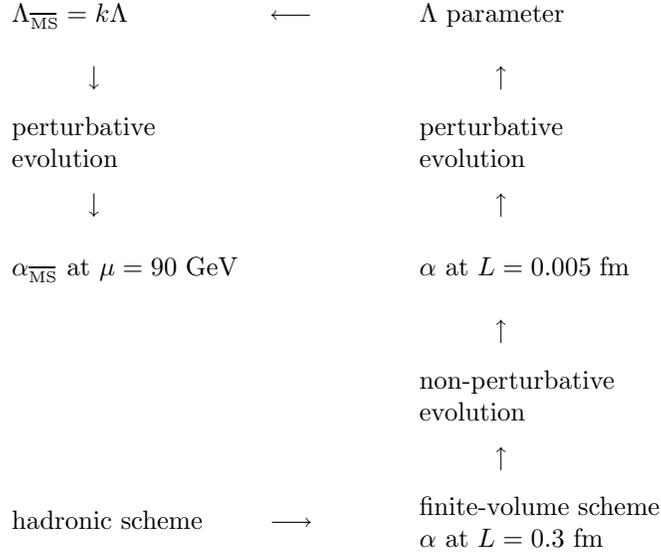

\begin{center}
\small
\renewcommand{\arraystretch}{2.0}
\begin{tabular}{l l c l}
\hbox{\kern2em}&$\Lambda_{\smallmsbar}=k\Lambda$
&\hspace{0.0cm}$\longleftarrow$\hspace{1.0cm} 
&$\Lambda$ parameter\\
&\hspace{1.0cm}$\downarrow$ 
& 
&\hspace{1.0cm}$\uparrow$ \\
&\begin{minipage}{3.0cm}
perturbative\\
evolution
\end{minipage} 
&&
\begin{minipage}{3.0cm}
perturbative\\
evolution
\end{minipage}\\
&\hspace{1.0cm}$\downarrow$ 
& 
&\hspace{1.0cm}$\uparrow$ \\
&$\alphaMSbar$ at $\mu=90$ GeV
&&
$\alpha$ at $L=0.005$ fm \\
&&
&\hspace{1.0cm}$\uparrow$ \\
&&&
\begin{minipage}{3.0cm}
non-perturbative\\
evolution
\end{minipage}\\
&&
&\hspace{1.0cm}$\uparrow$ \\
&hadronic scheme
&\hspace{0.0cm}$\longrightarrow$\hspace{1.0cm}
&\begin{minipage}{4.0cm}
finite-volume scheme\\
$\alpha$ at $L=0.3$ fm
\end{minipage}\\
\end{tabular}
\renewcommand{\arraystretch}{1.0}
\end{center}
\caption{Strategy to compute the running coupling,
taking low-energy data as input and using the non-perturbative
renormalization group to scale up to high energies.}
\label{Strategy}
\end{figure}

The computation of the running coupling discussed in this section 
follows the arrows in the diagram shown in 
fig.~\ref{Strategy}, starting at the lower-left corner.
In this plot the energy is increasing from the bottom to the top
while the entries in the left and right columns refer
to infinite and finite volume quantities respectively.

In the first step the chosen hadronic scheme is matched 
at low energies with a suitable finite-volume renormalization scheme.
The running coupling in this scheme is then evolved
non-perturbatively to higher energies using a recursive procedure.
Eventually the perturbative regime
is reached and the $\Lambda$ parameter can be determined
with negligible systematic uncertainty.
The conversion to the $\msbar$ scheme is trivial at this point
(top line of fig.~\ref{Strategy}). 

\begin{figure}[t]
\vspace{0.0cm}
\centerline{\epsfig{file=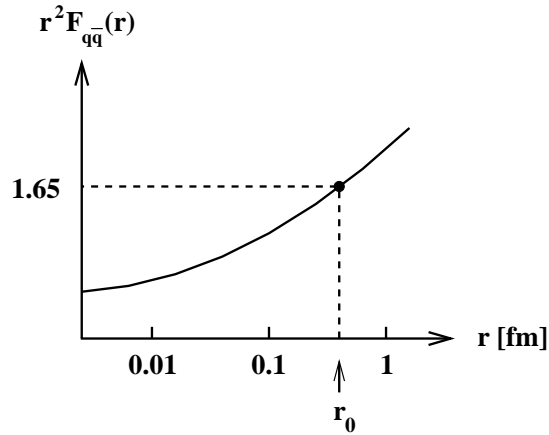,width=7.2cm}}
\vspace{0.0cm}
\caption{Definition of Sommer's scale $r_0$.
The dimensionless function $r^2\Fqqbar(r)$ is monotonically rising
and $r_0$ is equal to the distance $r$ where it passes through 1.65. 
\label{FigRnull}
}
\end{figure}

\subsection{Hadronic scheme}

Since we are considering the pure gauge theory, there are no
quark mass parameters to be fixed and a hadronic scheme is 
defined simply by specifying a low-energy reference scale.
We might take the mass of the lightest glueball, for example,
but in practice this would not be a good choice, 
because glueball masses are difficult to compute accurately
through numerical simulations.

The force $\Fqqbar(r)$ between static quarks at distance $r$
is more accessible in this respect 
and may be used to define a reference scale
$r_0$ through [\Cite{Rnulla}]
\begin{equation}
  r_0^2\Fqqbar(r_0)=1.65
  \label{DefRnull}
\end{equation} 
(see fig.~\ref{FigRnull}).
Comparing with the phenomenological charmonium potentials, 
the number on the right-hand side of eq.~(\ref{DefRnull}) has
been chosen so that $r_0\simeq0.5$ fm.
One should however keep in mind that the pure gauge theory
is unphysical and an assignment of physical units to $r_0$
should not be given too much significance. 

\exercise{At distances $r\geq0.3$ fm, the available simulation data 
for the heavy quark potential are well represented through
$V_{\rm q\bar{q}}(r)=\sigma r-{\pi\over12r}+c$. 
Calculate the string tension $\sigma$ assuming
$r_0=0.5$ fm}

\subsection{Finite-volume scheme}

We now need to specify the boundary conditions in finite volume and
to define a renormalized coupling
that scales with $L$.
For a number of technical reasons,
the coupling $\alphaqqbar$ introduced previously
is not employed, but in principle this would be an acceptable choice.
The definition given below is based on the Schr\"odinger functional
and requires some preparation.

Let us consider a lattice 
with boundary conditions as specified in section~5 (and no quark fields). 
Although we have not done so, it is possible to allow for 
arbitrary space-dependent boundary values and the associated 
partition function 
\begin{displaymath}
  {\cal Z}[C',C]=\int_{\rm fields}\rme^{-S}
\end{displaymath}
is then referred to as the Schr\"odinger functional.
Using the transfer matrix formalism, it is straightforward to show
that ${\cal Z}[C',C]$ is equal to the quantum mechanical 
transition amplitude for going from the field configuration $C$
at time $x_0=0$ to the configuration $C'$ at time $x_0=T$\/
[\Cite{Schrodingera}--\Cite{Schrodingerg}].

At small couplings the functional integral is dominated by the 
minimal action configuration $U(x,\mu)=\exp\{aB_{\mu}(x)\}$
with the specified boundary values.
In particular, the perturbation expansion of the Schr\"odinger functional,
\begin{equation}
  \Gamma[B]\equiv
  -\ln{\cal Z}[C',C]=
  {1\over g_0^2}\Gamma_0[B]+\Gamma_1[B]+g_0^2\Gamma_2[B]+\ldots,
  \label{SchrodingerExp}
\end{equation}
is obtained by expanding about this field, which in many respects
plays the r\^ole of a background field.
In eq.~(\ref{SchrodingerExp})
the leading term is equal to 
the classical action of the background field, 
while the higher-order terms require the calculation of Feynman diagrams
with propagators and vertices depending on $B$.

The renormalizability of the Schr\"odinger functional has been 
studied in perturbation theory and it turns out 
that all ultra-violet divergencies can be 
cancelled by including the usual counterterms 
plus a few boundary counterterms in the 
action~[\Cite{Schrodingerb}].
In pure gauge theories boundary counterterms in fact do not occur,
because there are no operators with the appropriate dimension and symmetries.
The Schr\"odinger functional
is hence a renormalized quantity in this case.
In particular, if we choose $C$ and $C'$ to 
depend on some parameter $\eta$, 
it is clear that a renormalized coupling
may be defined through
[\Cite{Schrodingerf}]
\begin{equation}
  \gbar^2=\left\{
  {\partial\Gamma_0\over\partial\eta}\biggm/
  {\partial\Gamma\over\partial\eta}
  \right\}_{\eta=0,T=L}.
\end{equation}
Note that $L$ is the only external scale in this formula
as it should be in a finite-volume renormalization scheme.

The precise choice of the boundary values is largely arbitrary
and was made essentially on the basis of 
practical considerations [\Cite{FSTb},\Cite{Schrodingerf}].
As in section~5 we take $C$ and $C'$ to be constant abelian
and set
\begin{eqnarray*}
  (\phi_1,\phi_2,\phi_3)&=&
  \frac{1}{3}\left(-1,0,1\right)\pi-
  \frac{1}{2}\left(-2,1,1\right)\eta,\\
  \noalign{\vskip1.5ex}
  (\phi'_1,\phi'_2,\phi'_3)&=&
  \frac{1}{3}\left(-3,1,2\right)\pi+
  \frac{1}{2}\left(-2,1,1\right)\eta.
\end{eqnarray*}
For small $\eta$ one can then prove that the background field
eq.~(\ref{bfieldb}) is the unique absolute minimum of the action
[\Cite{Schrodingerf}].

The calculation of $\gbar^2$ through numerical simulations 
does not present any particular problem. The important point
to note is that 
\begin{displaymath}
  \partial\Gamma/\partial\eta=
  \langle\partial S/\partial\eta\rangle
\end{displaymath}
is an expectation value of some combination of the gauge field
variables close to the boundaries.
Since gluon zero modes are excluded through the boundary conditions,
the commonly used simulation algorithms remain effective even
when $L$ is very small in physical units.

\exercise{Show that the background field eq.~(\ref{bfieldb})
solves the lattice field equations. Are there any other gauge
inequivalent solutions with the same boundary values?}

\subsection{Matching at low energies}

Returning to the diagram in fig.~\ref{Strategy},
we first need to match the hadronic scheme with the chosen finite-volume
scheme. In the present case 
this amounts to calculating the coupling $\gbar^2$ 
at some box size $L$ given in units of $r_0$. 
Evidently the matching should be done at some convenient point,
where both $r_0$ and $\gbar^2$ can be accurately 
computed using numerical simulations.
Preliminary calculations show that values of $\gbar^2$ around $3.5$
are in this range. For definiteness we take 
\begin{displaymath}
  \lmax\equiv\hbox{box size where $\gbar^2=3.480$}
\end{displaymath}
as the matching point and our task then is to determine the 
ratio $\lmax/r_0$.

To this end let us choose some value of the bare coupling $g_0$
where one is able to perform numerical simulations of physically
large lattices. Using standard techniques, 
$r_0/a$ may then be computed with negligible 
finite-volume errors [\Cite{Rnulla}--\Cite{Rnulle}].
At the same coupling one can also simulate lattices 
with smaller sizes, say $L/a=6,8,\ldots,16$, and calculate $\gbar^2$
in each case. The renormalized coupling is monotonically 
rising with $L/a$ and
$\lmax/a$ may hence be determined through interpolation 
to the point where $\gbar^2=3.480$ (if it is contained in 
the range of lattice sizes considered).

\begin{figure}[t]
\vspace{0.0cm}
\centerline{\epsfig{file=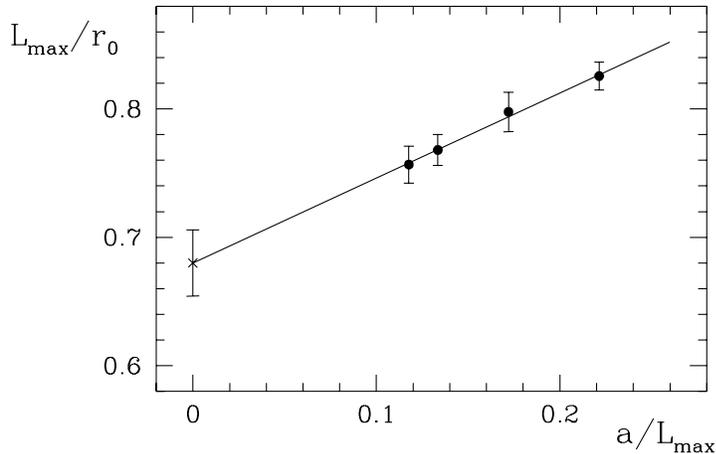,width=12.0cm}}
\vspace{0.0cm}
\caption{Simulation results for $\lmax/r_0$ 
and linear extrapolation to the continuum limit (left-most point).
\label{Lmax}
}
\end{figure}

Proceeding in this way the ratio
\begin{displaymath}
  \lmax/r_0=(\lmax/a)(r_0/a)^{-1}
\end{displaymath}
has been calculated at several couplings.
Different values of $g_0$ correspond to different lattice spacings
and any variation in the results that obtains 
thus signals the presence of lattice effects.
As shown in fig.~\ref{Lmax} the available data decrease
linearly with the lattice spacing and after extrapolation 
to the continuum limit one gets
\begin{equation}
  \lmax/r_0=0.680(26).
  \label{lmaxtornull}
\end{equation}
$\lmax$ is thus approximately equal to $0.34$ fm in physical units.
Note that the observed linear dependence on the lattice spacing 
is theoretically expected, even though there are no quarks here, 
because of the presence of the space-time boundaries
(see refs.~[\Cite{Schrodingerf},\Cite{paperI}] for further details).

\subsection{Non-perturbative renormalization group}

We now proceed to scale the renormalized coupling $\gbar^2$ to 
high energies (small box sizes), taking its value at $L=\lmax$
as initial datum. This will be achieved through a stepwise
procedure in which $L$ is decreased successively by factors of $2$.

Let us first consider 
the continuum theory, where the scale evolution of the coupling
is governed by the renormalization group equation
\begin{equation}
  L{\partial\gbar\over\partial L}=-\beta(\gbar)
  =b_0\gbar^3+b_1\gbar^5+\ldots,
  \label{RGequation}
\end{equation}
with $b_0=11(4\pi)^{-2}$ and $b_1=102(4\pi)^{-4}$ being the usual
coefficients. This equation determines the coupling at any scale 
if it is known at some point. In particular, there exists
a well-defined function $\sigma(u)$ such that
\begin{displaymath}
  \gbar^2(2L)=\sigma(\gbar^2(L)).
\end{displaymath}
$\sigma(u)$ may be regarded as 
an integrated form of the $\beta$-function and is referred to 
as the step scaling function.

\begin{figure}[t]
\vspace{0.0cm}
\centerline{\epsfig{file=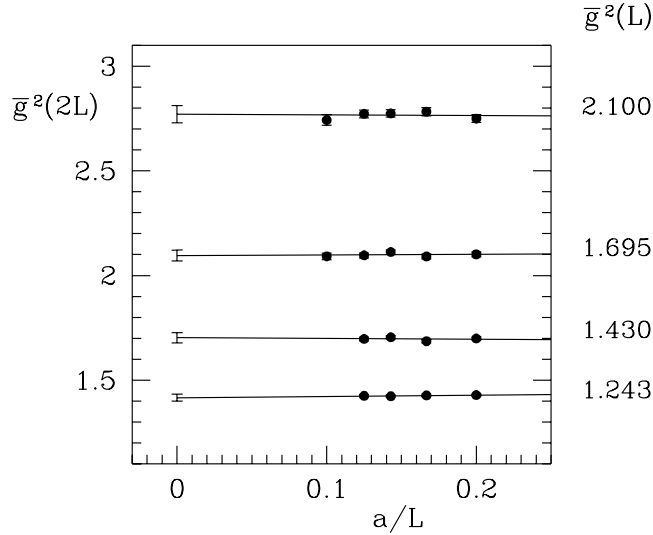,width=10.0cm}}
\vspace{-0.4cm}
\caption{Extrapolation of $\gbar^2(2L)$ to the continuum limit 
at various values of $\gbar^2(L)$.}
\label{SigmaExt}
\end{figure}

The crucial observation now is that $\sigma(u)$ can be computed
on the lattice through numerical simulations.
One first chooses some convenient lattice size,
say $L/a=8$, and adjusts the bare coupling so that 
$\gbar^2(L)=u$.
In practice the tuning which is required here does not cause any 
problems. We may then increase the lattice size by a factor $2$
and calculate $\gbar^2(2L)$ at the same value of the bare coupling.
The result of this computation is equal to $\sigma(u)$ up to 
lattice effects of order $a$ which can be extrapolated away
by repeating the calculation for several values of $L/a$.
As shown in fig.~\ref{SigmaExt} the simulation data are in fact 
independent of the lattice spacing within errors and 
the continuum limit can thus be easily reached.

\begin{figure}[t]
\vspace{0.0cm}
\centerline{\epsfig{file=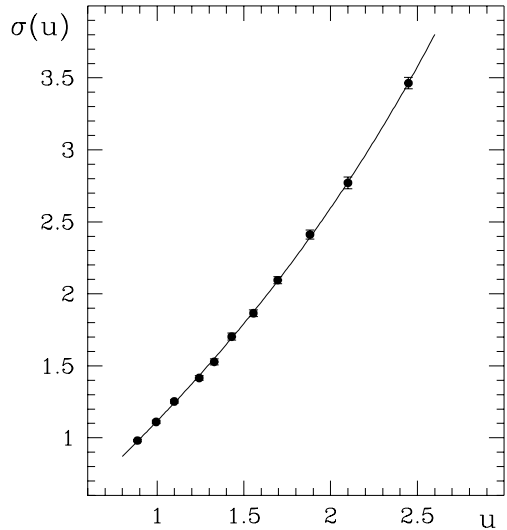,width=8.0cm}}
\vspace{-0.4cm}
\caption{Simulation results for the step scaling function
[\Cite{FSTb},\Cite{FSTe}].
The curve is a polynomial fit with two parameters as described in 
the text.}
\label{SigmaFit}
\end{figure}

Proceeding in this way the step scaling function has been
determined for a large range of couplings
(see fig.~\ref{SigmaFit}). At the smaller couplings the data compare well
with the expansion
\begin{equation}
  \sigma(u)=u+s_0u^2+s_1u^3+s_2u^4+\ldots,
  \label{SigmaPert}
\end{equation}
which one derives from eq.~(\ref{RGequation}),
the first two coefficients being
\begin{displaymath}
  s_0=2\ln(2)b_0,\qquad s_1=2\ln(2)b_1+s_0^2.
\end{displaymath}
We may in fact fit all the data by truncating the series 
at some low order and using the coefficients $s_2,s_3,\ldots$ as
fit parameters.

Once the step scaling function is known, the evolution of the coupling
can be computed straightforwardly by solving the recursion
\begin{displaymath}
  u_0=3.480,\qquad u_k=\sigma(u_{k+1}),
  \qquad k=0,1,2,\ldots
\end{displaymath}
The initial coupling $u_0$ corresponds $L=\lmax$ and 
one thus obtains the values of $\gbar^2$ at 
$L=2^{-k}\lmax$. If we demand that $\sigma(u)$ should 
only be used in the range of couplings covered by the data, 
the recursion can be applied $8$ times at most.
After six steps, for example, one gets
\begin{displaymath}
  \gbar^2=1.053(12),\qquad L=2^{-6}\lmax,
\end{displaymath}
where the error has been calculated by 
propagating the statistical errors to the fit polynomial
and solving the recursion using this function. 
Careful studies have been made to check that this 
gives correct error bounds and that the results do not depend on 
the number of fit parameters once a good fit quality is
achieved.

\exercise{Derive a recursion relation for the coefficients $s_k$
in eq.~(\ref{SigmaPert}) assuming the coefficients
$b_k$ of the $\beta$-function are known}

\subsection{Computation of the $\Lambda$ parameter}

\begin{figure}[t]
\vspace{0.0cm}
\centerline{\epsfig{file=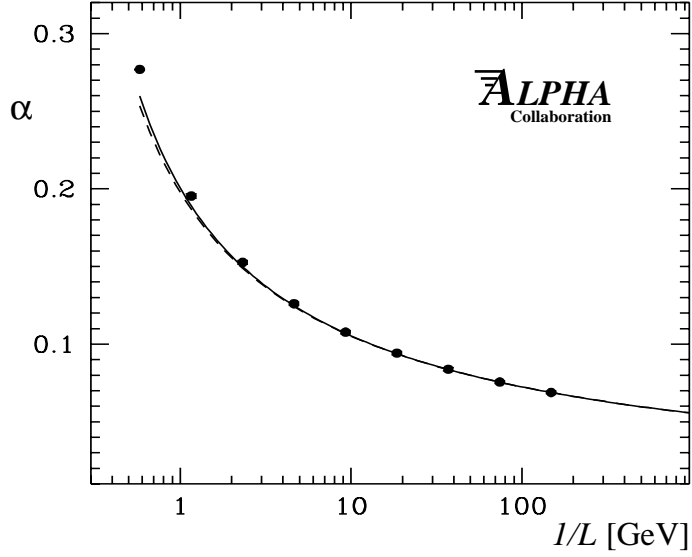,width=9.9cm}}
\vspace{-0.4cm}
\caption{Comparison of simulation results [\Cite{FSTb},\Cite{FSTe}]
for the running coupling
(data points) with perturbation theory. 
The solid (dashed) curve is obtained by integrating eq.~(\ref{RGequation}),
starting at the right-most point and using the 3-loop (2-loop)
$\beta$-function.}
\label{AlphaPlot}
\end{figure}

As shown in fig.~\ref{AlphaPlot} the data for the 
coupling $\alpha=\gbar^2/4\pi$ obtained in this way
cover a large range of energies from approximately
$600$ MeV to $150$ GeV (using $\lmax=0.34$ fm
to convert to physical units).
The perturbative scaling of the coupling sets in rather early
and for $\alpha\leq0.08$ the data lie on top of the 
two- and three-loop curves.
This may be a bit surprising, but one should not conclude
that the absence of large corrections to the perturbative
evolution is a general feature of the theory. 
In other schemes the coupling behaves differently
in general and there is usually no way to tell in advance
at which energy the non-perturbative contributions become
small.

We may now attempt to determine the $\Lambda$ parameter
by scaling the coupling all the way up to infinite energies 
and taking the limit
\begin{equation}
  \Lambda=\lim_{L\to0}
  {1\over L}\lambda(\gbar^2),
  \qquad
  \lambda(u)=(b_0u)^{-b_1/2b_0^2}\rme^{-1/2b_0u}.
  \label{LambdaAs}
\end{equation}
This is actually not the best way to proceed, because
the statistical and systematic errors would not be easy to trace.
A better starting point is the exact expression
\begin{equation}
  \Lambda=
  {1\over L}\lambda(\gbar^2)
  \exp\left\{\frac{1}{2}\int_0^{\gbar^2}\rmd u
  \left[{1\over b(u)}-{1\over b_0u^2}+{b_1\over b_0^2u}\right]\right\},
  \label{LambdaEx}
\end{equation}
where $b(u)$ is related to the $\beta$-function through
\begin{displaymath}
  b(\gbar^2)=-\gbar\beta(\gbar)=b_0\gbar^4+b_1\gbar^6+\ldots
\end{displaymath}
Eq.~(\ref{LambdaEx}) may be proved straightforwardly
by noting that the right-hand
side is independent of $L$ and that it has the correct behaviour 
for $L\to0$. 

To compute the $\Lambda$ parameter through this formula, one selects some 
value of $L$, say $L=2^{-6}\lmax$, where the coupling is known accurately 
and where it is sufficiently small that the perturbation expansion
for the $b$-function can be trusted. 
The series has recently been worked out 
to three loops [\Cite{Bode},\Cite{BodeWolffWeisz}]
and using this result to calculate the integral on the 
right-hand side of eq.~(\ref{LambdaEx}) one obtains
\begin{equation}
  \Lambda=0.211(16)/\lmax.
  \label{LambdaRes}
\end{equation}
Since the scale evolution of the coupling is accurately reproduced
by perturbation theory at high energies,
the systematic error arising from the 
neglected four-loop (and higher-order) corrections to the 
$b$-function are expected to be small.
Changing $L$ from $2^{-6}\lmax$ to $2^{-8}\lmax$ in fact 
does not have any appreciable effect on the calculated value of the $\Lambda$
parameter.
The influence of a hypothetical four-loop correction
can also be estimated by treating this term as a perturbation in 
eq.~(\ref{LambdaEx}) and 
assuming an approximately geometrical
progression of the expansion coefficients $b_k$.
The error bounds that one gets in this way 
are far below the statistical error and one again concludes
that the higher-order contributions can be neglected.

\exercise{Estimate the systematic uncertainty 
on the $\Lambda$ parameter which would arise 
if the three-loop coefficient $b_2$ would not be known}

\subsection{Conversion to the $\msbar$ scheme}

To one-loop order of perturbation theory the relation between
the finite-volume and the $\msbar$ coupling
is given by [\Cite{FSTb}]
\begin{displaymath}
  \alpha=\alphaMSbar(\mu)+
  \left[(11/2\pi)\ln(\mu L)-1.2556\right]\alphaMSbar(\mu)^2+\ldots
\end{displaymath}
A short calculation, starting from 
the definition eq.~(\ref{LambdaAs}),
then leads to the (exact) conversion factor $\LambdaMSbar/\Lambda=2.049$
and if this result is combined with eqs.~(\ref{lmaxtornull})
and (\ref{LambdaRes}) one ends up with
\begin{equation}
  \LambdaMSbar=0.636(54)/r_0.
  \label{LambdaMSbarRes}
\end{equation} 
Note that all reference to the intermediate finite-volume scheme has
disappeared at this point. Eq.~(\ref{LambdaMSbarRes}) simply
expresses the $\Lambda$ parameter in the $\msbar$ scheme in terms
of the low-energy scale $r_0$ and thus 
provides the solution of the non-perturbative renormalization problem
for the running coupling. 
In physical units, using $r_0=0.5$ fm,
one gets $\LambdaMSbar=251(21)$ MeV, but 
one should not forget that the conversion to physical units
is ambiguous in the pure gauge theory.
The solid result is eq.~(\ref{LambdaMSbarRes}).

The running coupling $\alphaMSbar(\mu)$ at normalization masses
$\mu$ given in units of $r_0$ may finally be calculated 
by solving the perturbative renormalization group equation 
in the $\msbar$ scheme, taking eq.~(\ref{LambdaMSbarRes}) as input. 
The numbers that one obtains tend to be significantly 
lower than the experimentally measured values of the 
strong coupling constant, but
since we have not included the quark polarization effects
there is no reason to be worried.
Sea quarks affect the evolution of the coupling (it becomes
flatter) and they also influence the low-energy
reference scale in some way which is difficult to foresee.

\section{Conclusions and outlook}

One of the greatest advantages of the lattice approach to QCD is 
that the theory is mathematically well-defined from the beginning.
The problem of non-perturbative renormalization, for example, 
can be given a precise meaning in this framework and 
its solution then reduces to a mainly computational task.
Ultimately the goal is to obtain a clean test of QCD by 
calculating several quantities with small statistical and systematic errors
and comparing the results with experiment.
This has not been achieved so far, but 
the theoretical advances described in these lectures
allow one to be more confident about the basic strategies.
In particular, as far as the continuum limit is concerned
it now seems unlikely that 
lattice spacings very much smaller than $0.1$ fm
will be needed for a reliable computation of the 
hadron masses and decay constants.

At this point the main problem in lattice QCD is that 
sea quark effects are not easily included in the 
numerical simulations. Although the available algorithms
have improved over the past few years, the amount of 
computer time required to simulate even a moderately large lattice
is still enormous. As a result of the rapid progress in computer technology, 
simulations of full QCD are however becoming increasingly feasible.
In the meantime the quenched approximation may be used as laboratory 
to test new ideas and to study more complicated 
physical problems ($K\to\pi\pi$ decays for example).

\end{document}